\documentclass[aps,prl,reprint,superscriptaddress]{revtex4-1}

\usepackage{graphicx}
\usepackage{siunitx}

\newcommand{\rucl}{$\alpha$-R\MakeLowercase{u}C\MakeLowercase{l}$_3$}

\begin{document}

\title{Finite field regime for a quantum spin liquid in $\alpha$-RuCl$_3$}

\author{Christian Balz}
\email{balzc@ornl.gov}
\affiliation{Neutron Scattering Division, Oak Ridge National Laboratory, Oak Ridge, Tennessee 37831, USA}
\author{Paula Lampen-Kelley}
\affiliation{Materials Science and Technology Division, Oak Ridge National Laboratory, Oak Ridge, Tennessee 37831, USA}
\affiliation{Department of Materials Science and Engineering, University of Tennessee, Knoxville, Tennessee 37996, USA}
\author{Arnab Banerjee}
\affiliation{Neutron Scattering Division, Oak Ridge National Laboratory, Oak Ridge, Tennessee 37831, USA}
\author{Jiaqiang Yan}
\affiliation{Materials Science and Technology Division, Oak Ridge National Laboratory, Oak Ridge, Tennessee 37831, USA}
\author{Zhilun Lu}
\affiliation{Helmholtz-Zentrum Berlin f\"ur Materialien und Energie, 14109 Berlin, Germany}
\author{Xinzhe Hu}
\author{Swapnil M. Yadav}
\author{Yasu Takano}
\affiliation{Department of Physics, University of Florida, Gainesville, Florida 32611, USA}
\author{Yaohua Liu}
\affiliation{Neutron Scattering Division, Oak Ridge National Laboratory, Oak Ridge, Tennessee 37831, USA}
\author{D. Alan Tennant}
\affiliation{Materials Science and Technology Division, Oak Ridge National Laboratory, Oak Ridge, Tennessee 37831, USA}
\author{Mark D. Lumsden}
\affiliation{Neutron Scattering Division, Oak Ridge National Laboratory, Oak Ridge, Tennessee 37831, USA}
\author{David Mandrus}
\affiliation{Materials Science and Technology Division, Oak Ridge National Laboratory, Oak Ridge, Tennessee 37831, USA}
\affiliation{Department of Materials Science and Engineering, University of Tennessee, Knoxville, Tennessee 37996, USA}
\author{Stephen E. Nagler}
\affiliation{Neutron Scattering Division, Oak Ridge National Laboratory, Oak Ridge, Tennessee 37831, USA}

\date{\today}

\begin{abstract}
An external magnetic field can induce a transition in {\rucl} from an ordered zigzag state to a disordered state that is possibly related to the Kitaev quantum spin liquid. Here we present new field dependent inelastic neutron scattering and magnetocaloric effect measurements implying the existence of an additional transition out of the quantum spin liquid phase at an upper field limit $B_u$.   The neutron scattering shows three distinct regimes of magnetic response.  In the low field ordered state the response shows magnon peaks; the intermediate field regime shows only continuum scattering, and above $B_u$ the response shows sharp magnon peaks at the lower bound of a strong continuum.   Measurable dispersion of magnon modes along the $(0,0,L)$ direction implies non-negligible inter-plane interactions.     Combining the magnetocaloric effect measurements with other data  a $T-B$ phase diagram is constructed. The results constrain the range where one might expect to observe quantum spin liquid behavior in {\rucl}.
\end{abstract}

\maketitle

The fractional Majorana fermion excitations of a Kitaev quantum spin liquid (QSL) \cite{Kit06} have been proposed as a route to topologically protected qubits \cite{Nay08}. The suggestion that this physics is exhibited in certain honeycomb magnets with $J_{\text{eff}} = 1/2$ ground states \cite{Jac09} led to an enormous amount of research on iridate materials \cite{Sin12,Cho12,Chu15,Rui17,Kit18}, and more recently, an intense interest in {\rucl} \cite{Plu14,Kub15,Joh15,Sea15}. In the absence of an external magnetic field {\rucl} orders below $T_N\approx 7$~K in a 3D stacked antiferromagnetic zigzag ground state \cite{Cao16,Par16}, however an in-plane field perpendicular to a Ru-Ru bond of $B_{c}\approx7.5$~T results in a quantum disordered phase proposed to be a QSL \cite{Yad16}. The magnetic excitations of {\rucl} have been studied using various techniques including Raman scattering \cite{San15}, ESR and THz spectroscopy \cite{Pon17,Wan17,Lit17,Wu18,Shi18,Wel18}, and inelastic neutron scattering (INS) \cite{Ban16,Ban17,Ran17,Do17}. Spectroscopic evidence for fractionalization is seen in the form of unusual continuum scattering around the 2D $\Gamma$ point \cite{San15,Wan17,Ban17,Do17}.  Above $B_c$ the spin waves associated with zigzag order disappear and the continuum is enhanced \cite{Ban18}. 

The report of a quantized thermal Hall plateau \cite{Kas18} in a finite field range above $B_c$ provided further evidence that the continuum excitations are related to chiral Majorana edge modes.  Various theoretical proposals have been advanced to explain these observations \cite{Ye18,Coo18,Vin18,Jia19,Gor19,Tak19}, however, to date the thermal Hall results have not been confirmed by other groups, and there has been no independent report of a suggested additional higher field topological transition out of the disordered regime.  Here we present new field dependent magnetocaloric effect (MCE) measurements showing clear thermodynamic evidence for a transition at an upper field $B_u$, at which INS shows a qualitative change in the magnetic excitation spectrum.  This has all of the characteristics of a topological transition, implying that the proposed QSL phase is distinct from and is not smoothly connected to a field polarized paramagnetic state.  The low temperature regime of the $T-B$ phase diagram is clarified, showing the $T$ dependence of $B_{c}$ and $B_{u}$.

We begin by presenting the neutron scattering data collected on a 2~g single crystal grown using vapor-transport techniques as described elsewhere \cite{Ban17}. The INS measurements utilized the FLEXX triple-axis spectrometer at the Helmholtz-Zentrum Berlin \cite{Hab15}. The spectrometer was configured with open collimation, double focusing PG monochromator, and horizontally focusing PG analyzer with a fixed final energy of 5~meV, yielding an energy resolution of 0.34~meV FWHM at the elastic position. The crystal was mounted initially with the $(H,0,L)$ scattering plane horizontal (using trigonal notation). In an applied field there was a slight \ang{6} irreversible rotation about the $L$ axis. Given the known dependence of $B_c$ on the precise in-plane field direction \cite{Lam18_2} this has no significant effect on the results shown below.

\begin{figure*}
\includegraphics[width=\textwidth]{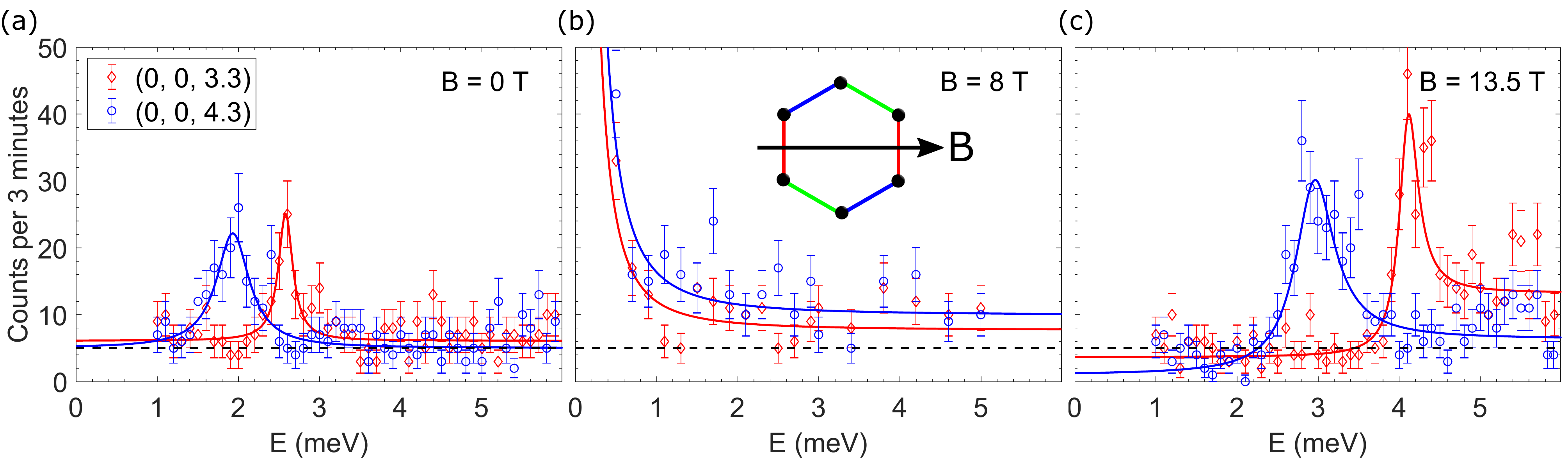}
\caption{Field dependence of the inelastic neutron scattering at the 2D $\Gamma$-point for two values of the out-of-plane wavevector transfer. Data obtained at 1.5~K on a 2~g single crystal of  {\rucl} using the FLEXX triple-axis spectrometer. (a) Zero-field data. A field of (b) 8~T and (c) 13.5~T was applied in the honeycomb plane perpendicular to a Ru-Ru bond (see inset of (b)). The solid lines are fits and the dashed lines show the model free background for $(0,0,3.3)$ as described in the text. Error bars represent one standard deviation assuming Poisson statistics.\label{FLEXX}}
\end{figure*}

Figure~\ref{FLEXX} shows constant wavevector scans for two values of $(0,0,L)$ corresponding to the 2D $\Gamma$-point. At energies below 1 meV incoherent elastic scattering dominates the signal. In the accessible inelastic range the data shows three distinct field regimes corresponding to (I)~$B<B_c$,  (II)~$B_c<B<B_u$, and (III)~$B>B_u$. In region I, including zero field (figure~\ref{FLEXX}(a)), well-defined magnon peaks are visible arising from the zigzag order. The peak energies depend on $L$ as expected for 3D zigzag magnetic order. Above $B_c$ in region II the magnon peaks disappear as reported previously \cite{Ban18}. In region III, above $B_u$, a sharp gapped magnon mode reappears. Additional modes may be present at energies above those measurable in the current experiment. The continuum scattering reported previously is present in regions I and III at energies above the magnon peaks as well as in region II. These results may be compared with previous THz measurements, albeit where the direction of the in-plane field was not specified \cite{Wan17}. They show magnon peaks at low and high fields, and although $B_u$ was not identified there is a region near 7~T where no such peaks are evident. In contrast, ESR measurements \cite{Pon17} have been interpreted as showing modes at all fields.

The statistical significance of the continuum scattering level in the INS can be assessed in a model free fashion using the average count levels over selected ranges of energy transfer. The effective background level is determined from the low energy scattering measured at 13.5~T as this represents the cleanest signal. The levels for the scans depicted in figure~\ref{FLEXX} are shown in table~\ref{counts}. Note that all time-of-flight (ToF) measurements reported previously \cite{Ban17,Do17,Ban18} for the 2D $\Gamma$ point in regions I and II represent integrations over a large range of $L$.  The ToF procedure captures more of the continuum scattering, and the measurements at $B=8$~T were interpreted as showing a gap \cite{Ban18}. In the present experiment the statistics measured at single wavevectors are insufficient to confirm the value of any possible gap in region II. The continuum intensity, at least in regions I and II,  does not show any significant dependence on $L$, consistent with the 2D nature of the continuum scattering reported previously \cite{Ban17}.

\begin{table}
\caption{Average counts per 3 minutes. The uncertainties represent one standard deviation of the last digit assuming Poisson statistics.\label{counts}}
\begin{ruledtabular}
\begin{tabular}{c|c c c c}
 & Range [meV] & Range [meV] & Counts & Counts \\
 & $L=3.3$ & $L=4.3$ & $L=3.3$ & $L=4.3$ \\
\hline
Background & $1-3.5$ & $1-2$ & 5.0(5) & 4.2(7) \\
0 T & $3-6$ & $2.5-6$ & 7.4(5) & 6.4(5)\\
8 T & $1-5$ & $1-5$ & 9.7(8) & 13(1)\\
13.5 T & $4.5-6$ & $4-6$ & 15(1) & 8.3(7)
\end{tabular}
\end{ruledtabular}
\end{table}

\begin{figure}
\includegraphics[width=\columnwidth]{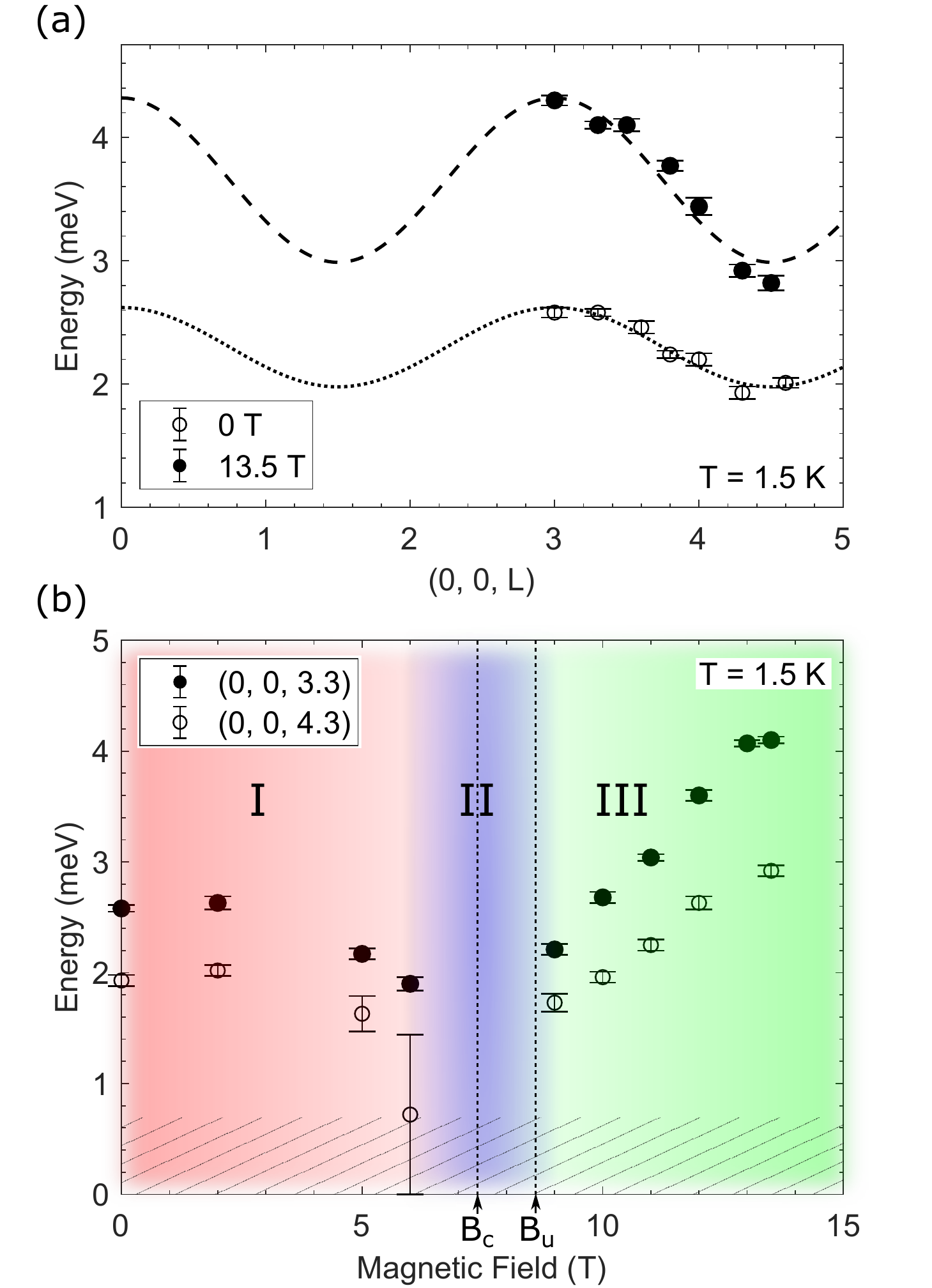}
\caption{(a) The $L$ dependence of the spin wave energy observed at the 2D $\Gamma$-point at 0~T and 13.5~T. Dashed and dotted lines are fits to the dispersion as described in the text. (b) The magnetic field dependence of the 2D $\Gamma$-point spin wave energy measured at wavevector transfers $(0,0,3.3)$ and $(0,0,4.3)$, close to the minimum and maximum of the $L$ dispersion. The colors indicate three regions of the inelastic response as described in the text. The hatched region marks the low energy range not resolved in this data. In region II no magnon peaks were identified in the neutron scattering data.  The region between $B_c$ and $B_u$ determined by MCE is shown by the vertical dashed lines.  In both panels the error bars represent one standard deviation of the fitted peak positions.\label{LB_dep}}
\end{figure}

To extract values of the magnon peak positions the data in regions I and III was fitted to an empirical function consisting of the sum of: a Lorentzian representing the peak, a constant background, and a rounded hyperbolic tangent function with the origin at the peak position to model the continuum scattering on the high energy side. The data in region II was fitted to a peak centered at zero energy plus a constant background. The fitted peak positions are used to plot the dispersion of the magnons along $(0,0,L)$ as shown in figure~\ref{LB_dep}(a) for zero field (open circles) and $B = 13.5$~T (closed circles).  The spectrum at 13.5 T is shifted upwards, presumably by the Zeeman energy, and has a larger bandwidth than that in zero field, but the periodicity of the dispersion is unchanged. The dispersion is clear evidence of out-of-plane magnetic interactions, consistent with the 3D stacked zigzag order in zero field. The excitations in a fully polarized ferromagnetic state are expected to be sharp magnons with the dispersion determined by the underlying interactions \cite{Col02}. The presence of the high energy continuum at $B = 13.5$~T is consistent with partial ferromagnetic polarization.    

Empirically, the dispersion of the mode along $L$ can be modeled by the simple function 
\begin{equation}
E=A+B\cos\left(\frac{2\pi L}{3}\right)
\end{equation}
where $A$ represents an average magnon energy along $(0,0,L)$ including the Zeeman term, and the bandwidth $B$ reflects interlayer coupling.  From considerations of the 3D magnetic order the effective interlayer coupling is expected to be antiferromagnetic \cite{Cao16,Par16}. A spin wave calculation can, in principle, lead to the observed periodicity along $(0,0,L)$ but the full 3D Hamiltonian for {\rucl} remains unresolved and there is insufficient information to present a meaningful 3D model here. The fact that the in-plane interactions in particular are expected to be heavily frustrated makes it difficult to estimate the relative magnitude of the in-plane and inter-plane couplings. A rough upper bound can be obtained by comparing the in-plane bandwidth of the lowest magnon mode with that of the dispersion along $L$.  For example, in a simple model with nearest-neighbor Heisenberg couplings, $J$, the leading order term of the out-of-plane dispersion is proportional to $\sqrt{JJ_\perp}$ \cite{Sat80} where $J_\perp$ is the out-of-plane coupling. The zero field out-of-plane magnon bandwidth for {\rucl} is 0.6 meV.  Experimental estimates of the in-plane bandwidth of the lowest spin wave mode vary from 1.3 meV \cite{Ban18} to 5.5 meV \cite{Ran17}, corresponding in a simple model to ratios of $J_\perp/J=1-20\%$. Frustration is expected to lead to a narrower in-plane bandwidth, and accounting for this we expect the inter-plane coupling in {\rucl} to be at most a few percent of the in-plane coupling.  An accurate determination of the full Hamiltonian requires at minimum a measurement of the in-plane magnon dispersion in the high field limit, a significant undertaking beyond the scope of this Letter.

The point $(0,0,0)$ sampled by THz spectroscopy and other optical techniques is a local maximum of the dispersion along $L$ and the excitations there do not represent an overall energy gap. The $L$ dispersion might explain the difference in the gap energy at 13.5 T inferred by thermal conductivity (2.8 meV \cite{Hen18}) vs. that seen in THz absorption (4.5 meV \cite{Wan17}) since the former is presumably related to the global excitation energy minimum and agrees with the INS results at $(0,0,4.5)$.  

The wavevectors ($L=3.3$ and $L=4.3$) plotted in figure~\ref{FLEXX} are near the maximum ($L$=3) and minimum ($L=4.5$) of the $L$-dispersion.  The peak positions as a function of magnetic field are plotted in figure~\ref{LB_dep}(b). Inelastic peaks below 0.7 meV (hatched region) were undetectable.  In region I the spin wave energies diminish with increasing field, disappearing completely at $B_c$, consistent with the ToF measurements \cite{Ban18}. No peaks could be discerned in region II and the scattering is dominated by the continuum, consistent with expectations for a QSL \cite{Kno14}. At 9~T and above (region III) sharp peaks reappear with a strong continuum on the high energy side. The sudden change in the spectrum is consistent with a value for $B_u$ between 8.5~T and 9~T. The peak energy increases roughly linearly with field in region III, and in the overlap region agrees with the gap energy extracted from thermal conductivity measurements \cite{Hen18}. The slope of the mode energies in the high field regime above $B_u$ is consistent with the results of other techniques \cite{Wan17,Wel18,Pon17}. As seen in figure~\ref{LB_dep}(b), in region III the bandwidth increases with increasing field. The narrower bandwidth just above $B_u$ is possibly an effect of quantum fluctuations and implies that as the QSL is approached the layers decouple, underscoring the 2D nature of the physics in region II.

The $T-B$ phase diagram was investigated further by MCE measurements spanning the region $0.8-7$~K and $5-11$~T. These used a 6~mg single crystal, with a sharp zero-field phase transition near 7~K as determined by heat capacity measurements. The magnetic field applied in the $ab$ plane was swept both upward and downward at a rate of 0.3~T/min. By comparison with the known angular dependence of the transition temperature at $B_c$ \cite{Lam18_2}, the direction of the field was identified as \ang{10} off from the direction perpendicular to a Ru-Ru bond.  The field sweep produces a temperature difference $\Delta T$ between the sample and the thermal reservoir that depends on $\left(\frac{\partial M}{\partial T}\right)_H$ \cite{Sch97} where $M$ is the magnetization. Reversing the field-sweep direction reverses the sign of the temperature difference and phase transitions are identified by anomalies in the difference between the curves.

\begin{figure}
\includegraphics[width=\columnwidth]{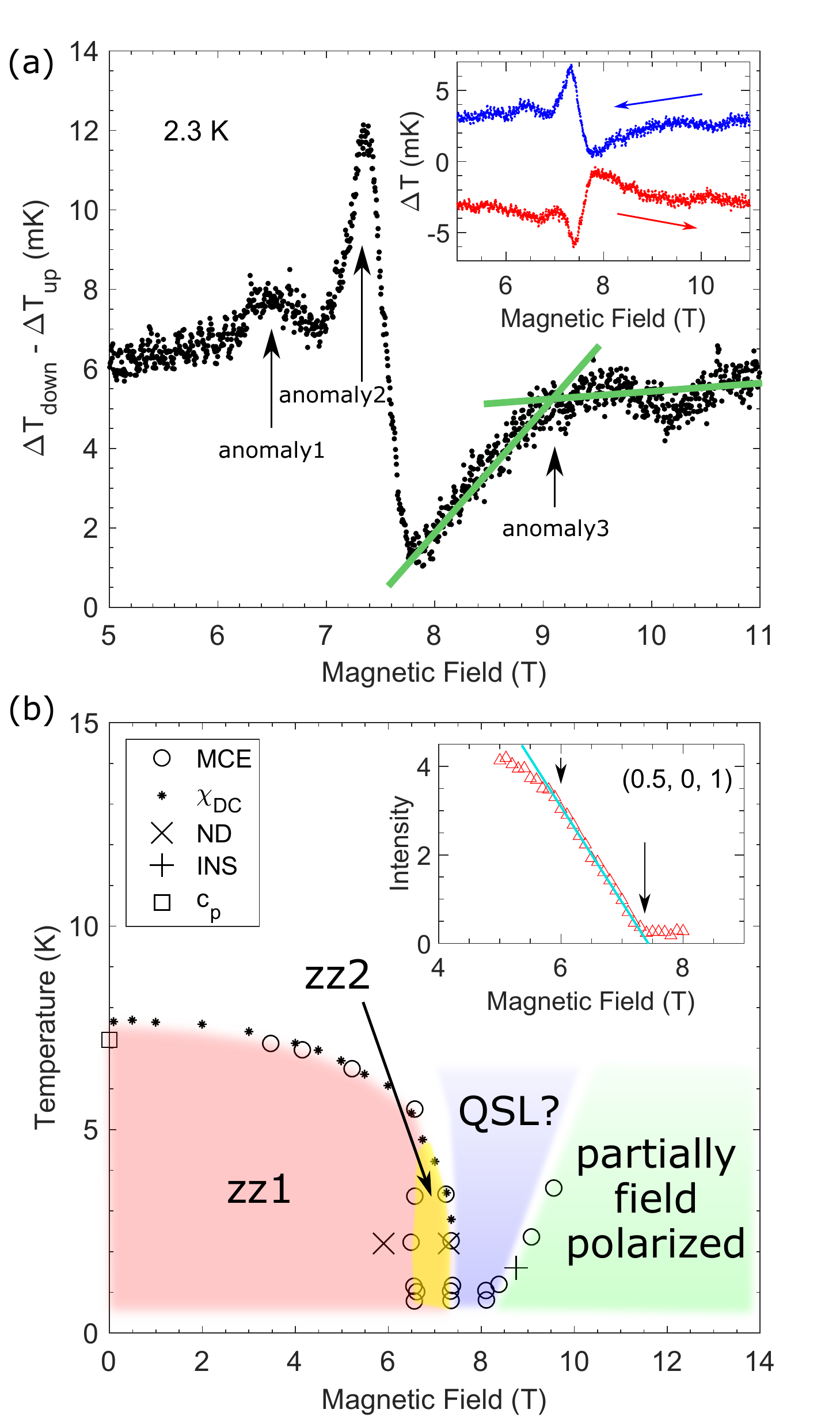}
\caption{(a) MCE difference curve at 2.3~K. The black arrows show the locations of transitions (see text). Inset: Temperature difference between sample and heat bath due to MCE.  The red and blue arrows indicate the field sweep directions.  (b) The $T-B$ phase diagram of {\rucl} as determined in this work. The phase boundaries were deduced from magnetocaloric effect (MCE), neutron diffraction (ND), inelastic neutron scattering (INS), specific heat ($c_p$), and magnetic susceptibility ($\chi_{DC}$) obtained previously \cite{Ban18}. Four potential phases are indicated by the colors. Inset: Magnetic Bragg peak intensity at 2.2~K as a function of magnetic field, measured using the CORELLI instrument at SNS. The solid line is a guide to the eye. \label{mce_pd}}
\end{figure}

The main panel of figure~\ref{mce_pd}(a) shows MCE difference data at 2.3~K, revealing three anomalies in the measured field range. Anomaly 1 (6.5 T at 2.3 K) appears as a weak maximum at a transition seen previously in AC susceptibility \cite{Lam18_2} corresponding to a change between two zigzag ordered states with different interlayer stacking \cite{Bal19}.  The sharp maximum at anomaly 2 (7.35 T at 2.3 K) is the transition at $B_c$. These two transitions are also evident in the intensity of magnetic Bragg peaks seen in neutron diffraction (ND, see arrows in the inset of figure~\ref{mce_pd}(b)) as the intensity changes slope at 5.9~T and disappears at 7.3~T. Anomaly 3 appears as a kink in the MCE difference and defines $B_u$. Additional details of the MCE measurements are found in the supplemental material \cite{SuppMat}.

A $T-B$ phase diagram assembled from the available data is shown in Figure~\ref{mce_pd}(b) utilizing the new MCE and previously published susceptibility data \cite{Ban18}. Additional points are obtained from the INS, the zero-field heat capacity, and ND.  The phase diagram shows four phases: two ordered states labeled zz1 and zz2 comprising region I, the potential QSL state (II), and a state that is partially field polarized (III). The transition points derived from the different measurements are mutually consistent except for a minor detail: the zz2 phase at 2.2~K seen in ND has a lower onset than that inferred from MCE.  This is likely because the ND was taken with the field precisely perpendicular to a Ru-Ru bond direction, and the width of the zz2 phase is known to depend on the in-plane field direction \cite{Lam18_2}. 

The sharp magnon seen above $B_{u}$ in INS indicates a distinct change in the ground state, and the fact that it is accompanied by a thermodynamic anomaly visible in the MCE is consistent with a topological transition as suggested from thermal Hall measurements \cite{Kas18}.  This implies the addition of a high field state to existing phase diagrams for {\rucl} \cite{Zhe17,Sea17,Wol17,Bae17,Ban18} and it is likely that there are additional transitions at fields above $B_u$, eventually leading to a fully polarized state. The latter would be indicated by a crossing of the two MCE curves shown in the inset of figure~\ref{mce_pd}(a) since $\left(\frac{\partial M}{\partial T}\right)_H$ is known to change sign at the saturation field (see e.g. \cite{For09}).

The difference between $B_c$ and $B_u$ evidently decreases with temperature between 4~K and 1~K. Whether or not the transition lines converge at $T=0$ remains unresolved experimentally. If in fact the lines converge at $T=0$ the fractional excitations seen at 1.5~K may be a signature of a quantum critical point (QCP) rather than a finite field region with a $T=0$ QSL ground state. Exact diagonalization calculations for finite systems using a Hamiltonian proposed to describe {\rucl} have been interpreted as implying that the quantum disordered state is smoothly connected to the field polarized state \cite{Win18}.  This might be consistent with such a QCP, and lower temperature measurements are called for to help resolve this issue.  

Many theoretical works have pointed to off-diagonal exchange (so called $\Gamma$ and $\Gamma'$ terms) in the Hamiltonian playing an important role in the possible field-induced (or field-revealed)  QSL behavior of {\rucl}, see e.g. \cite{Yad16,McC18,Gor19,Tak19,Jia19}. It has been argued recently \cite{Tak19} that in the presence of an external field such a term leads to a mass gap in the Majorana fermion spectrum, a necessary condition for observing quantization in the thermal Hall effect.  Most of these calculations consider a field applied in the $<111>$ direction in \emph{spin space}, corresponding to a field perpendicular to the honeycomb plane. New density matrix renormalization group calculations \cite{Gor19} exploring the effect of field direction show that the region of QSL behavior between the zigzag ordered and fully polarized phase is largest for an external field in this direction, but even with the field applied solely in-plane a narrow intermediate QSL region is possible.  The field required to destroy the zigzag order quickly becomes very large when rotated out of the honeycomb plane. 

In summary, the measurements reported here show clear evidence for field induced transitions in {\rucl}, including a probable topological transition from a quantum disordered phase to a partially polarized phase in agreement with the suggestion arising from thermal Hall effect measurements \cite{Kas18}. INS measurements of magnetic excitations at wavevectors corresponding to the 2D $\Gamma$ point do not show any evidence of sharp magnons in the region of the phase diagram that might be associated with a QSL. Looking to the future, an unambiguous determination of the effective spin Hamiltonian for {\rucl} calls for INS measurements of excitations in the polarized state over the entire Brillouin zone. Equally, a full theoretical description of the magnetic transitions in {\rucl} will need to include inter-plane interactions. The present results for the $T-B$ phase diagram provide important constraints for such a theory.

\begin{acknowledgments}
We thank Mark Meisel and Steven Kivelson for valuable discussions. P.L-K. and D. M. acknowledge support from the Gordon and Betty Moore Foundation’s EPiQS Initiative Grant No. GBMF4416. We acknowledge support from the U.S. Department of Energy (U.S.-DOE), Office of Science - Basic Energy Sciences (BES), Materials Sciences and Engineering Division. Work at the Oak Ridge National Laboratory Spallation Neutron Source was supported by U.S.-DOE, Office of Science - BES, Scientific User Facilities Division. We thank the Helmholtz-Zentrum Berlin for the allocation of neutron beamtime. A portion of this work was performed at the National High Magnetic Field Laboratory, which is supported by the National Science Foundation Cooperative Agreement No. DMR-1644779 and the State of Florida.
\end{acknowledgments}


\begin{thebibliography}{50}%
\makeatletter
\providecommand \@ifxundefined [1]{%
 \@ifx{#1\undefined}
}%
\providecommand \@ifnum [1]{%
 \ifnum #1\expandafter \@firstoftwo
 \else \expandafter \@secondoftwo
 \fi
}%
\providecommand \@ifx [1]{%
 \ifx #1\expandafter \@firstoftwo
 \else \expandafter \@secondoftwo
 \fi
}%
\providecommand \natexlab [1]{#1}%
\providecommand \enquote  [1]{``#1''}%
\providecommand \bibnamefont  [1]{#1}%
\providecommand \bibfnamefont [1]{#1}%
\providecommand \citenamefont [1]{#1}%
\providecommand \href@noop [0]{\@secondoftwo}%
\providecommand \href [0]{\begingroup \@sanitize@url \@href}%
\providecommand \@href[1]{\@@startlink{#1}\@@href}%
\providecommand \@@href[1]{\endgroup#1\@@endlink}%
\providecommand \@sanitize@url [0]{\catcode `\\12\catcode `\$12\catcode
  `\&12\catcode `\#12\catcode `\^12\catcode `\_12\catcode `\%12\relax}%
\providecommand \@@startlink[1]{}%
\providecommand \@@endlink[0]{}%
\providecommand \url  [0]{\begingroup\@sanitize@url \@url }%
\providecommand \@url [1]{\endgroup\@href {#1}{\urlprefix }}%
\providecommand \urlprefix  [0]{URL }%
\providecommand \Eprint [0]{\href }%
\providecommand \doibase [0]{http://dx.doi.org/}%
\providecommand \selectlanguage [0]{\@gobble}%
\providecommand \bibinfo  [0]{\@secondoftwo}%
\providecommand \bibfield  [0]{\@secondoftwo}%
\providecommand \translation [1]{[#1]}%
\providecommand \BibitemOpen [0]{}%
\providecommand \bibitemStop [0]{}%
\providecommand \bibitemNoStop [0]{.\EOS\space}%
\providecommand \EOS [0]{\spacefactor3000\relax}%
\providecommand \BibitemShut  [1]{\csname bibitem#1\endcsname}%
\let\auto@bib@innerbib\@empty
\bibitem [{\citenamefont {Kitaev}(2006)}]{Kit06}%
  \BibitemOpen
  \bibfield  {author} {\bibinfo {author} {\bibfnamefont {A.}~\bibnamefont
  {Kitaev}},\ }\href@noop {} {\bibfield  {journal} {\bibinfo  {journal} {Annals
  of Physics}\ }\textbf {\bibinfo {volume} {321}},\ \bibinfo {pages} {2}
  (\bibinfo {year} {2006})}\BibitemShut {NoStop}%
\bibitem [{\citenamefont {Nayak}\ \emph {et~al.}(2008)\citenamefont {Nayak},
  \citenamefont {Simon}, \citenamefont {Stern}, \citenamefont {Freedman},\ and\
  \citenamefont {Das~Sarma}}]{Nay08}%
  \BibitemOpen
  \bibfield  {author} {\bibinfo {author} {\bibfnamefont {C.}~\bibnamefont
  {Nayak}}, \bibinfo {author} {\bibfnamefont {S.~H.}\ \bibnamefont {Simon}},
  \bibinfo {author} {\bibfnamefont {A.}~\bibnamefont {Stern}}, \bibinfo
  {author} {\bibfnamefont {M.}~\bibnamefont {Freedman}}, \ and\ \bibinfo
  {author} {\bibfnamefont {S.}~\bibnamefont {Das~Sarma}},\ }\href {\doibase
  10.1103/RevModPhys.80.1083} {\bibfield  {journal} {\bibinfo  {journal} {Rev.
  Mod. Phys.}\ }\textbf {\bibinfo {volume} {80}},\ \bibinfo {pages} {1083}
  (\bibinfo {year} {2008})}\BibitemShut {NoStop}%
\bibitem [{\citenamefont {Jackeli}\ and\ \citenamefont
  {Khaliullin}(2009)}]{Jac09}%
  \BibitemOpen
  \bibfield  {author} {\bibinfo {author} {\bibfnamefont {G.}~\bibnamefont
  {Jackeli}}\ and\ \bibinfo {author} {\bibfnamefont {G.}~\bibnamefont
  {Khaliullin}},\ }\href {\doibase 10.1103/PhysRevLett.102.017205} {\bibfield
  {journal} {\bibinfo  {journal} {Phys. Rev. Lett.}\ }\textbf {\bibinfo
  {volume} {102}},\ \bibinfo {pages} {017205} (\bibinfo {year}
  {2009})}\BibitemShut {NoStop}%
\bibitem [{\citenamefont {Singh}\ \emph {et~al.}(2012)\citenamefont {Singh},
  \citenamefont {Manni}, \citenamefont {Reuther}, \citenamefont {Berlijn},
  \citenamefont {Thomale}, \citenamefont {Ku}, \citenamefont {Trebst},\ and\
  \citenamefont {Gegenwart}}]{Sin12}%
  \BibitemOpen
  \bibfield  {author} {\bibinfo {author} {\bibfnamefont {Y.}~\bibnamefont
  {Singh}}, \bibinfo {author} {\bibfnamefont {S.}~\bibnamefont {Manni}},
  \bibinfo {author} {\bibfnamefont {J.}~\bibnamefont {Reuther}}, \bibinfo
  {author} {\bibfnamefont {T.}~\bibnamefont {Berlijn}}, \bibinfo {author}
  {\bibfnamefont {R.}~\bibnamefont {Thomale}}, \bibinfo {author} {\bibfnamefont
  {W.}~\bibnamefont {Ku}}, \bibinfo {author} {\bibfnamefont {S.}~\bibnamefont
  {Trebst}}, \ and\ \bibinfo {author} {\bibfnamefont {P.}~\bibnamefont
  {Gegenwart}},\ }\href {\doibase 10.1103/PhysRevLett.108.127203} {\bibfield
  {journal} {\bibinfo  {journal} {Phys. Rev. Lett.}\ }\textbf {\bibinfo
  {volume} {108}},\ \bibinfo {pages} {127203} (\bibinfo {year}
  {2012})}\BibitemShut {NoStop}%
\bibitem [{\citenamefont {Choi}\ \emph {et~al.}(2012)\citenamefont {Choi},
  \citenamefont {Coldea}, \citenamefont {Kolmogorov}, \citenamefont
  {Lancaster}, \citenamefont {Mazin}, \citenamefont {Blundell}, \citenamefont
  {Radaelli}, \citenamefont {Singh}, \citenamefont {Gegenwart}, \citenamefont
  {Choi}, \citenamefont {Cheong}, \citenamefont {Baker}, \citenamefont
  {Stock},\ and\ \citenamefont {Taylor}}]{Cho12}%
  \BibitemOpen
  \bibfield  {author} {\bibinfo {author} {\bibfnamefont {S.~K.}\ \bibnamefont
  {Choi}}, \bibinfo {author} {\bibfnamefont {R.}~\bibnamefont {Coldea}},
  \bibinfo {author} {\bibfnamefont {A.~N.}\ \bibnamefont {Kolmogorov}},
  \bibinfo {author} {\bibfnamefont {T.}~\bibnamefont {Lancaster}}, \bibinfo
  {author} {\bibfnamefont {I.~I.}\ \bibnamefont {Mazin}}, \bibinfo {author}
  {\bibfnamefont {S.~J.}\ \bibnamefont {Blundell}}, \bibinfo {author}
  {\bibfnamefont {P.~G.}\ \bibnamefont {Radaelli}}, \bibinfo {author}
  {\bibfnamefont {Y.}~\bibnamefont {Singh}}, \bibinfo {author} {\bibfnamefont
  {P.}~\bibnamefont {Gegenwart}}, \bibinfo {author} {\bibfnamefont {K.~R.}\
  \bibnamefont {Choi}}, \bibinfo {author} {\bibfnamefont {S.-W.}\ \bibnamefont
  {Cheong}}, \bibinfo {author} {\bibfnamefont {P.~J.}\ \bibnamefont {Baker}},
  \bibinfo {author} {\bibfnamefont {C.}~\bibnamefont {Stock}}, \ and\ \bibinfo
  {author} {\bibfnamefont {J.}~\bibnamefont {Taylor}},\ }\href {\doibase
  10.1103/PhysRevLett.108.127204} {\bibfield  {journal} {\bibinfo  {journal}
  {Phys. Rev. Lett.}\ }\textbf {\bibinfo {volume} {108}},\ \bibinfo {pages}
  {127204} (\bibinfo {year} {2012})}\BibitemShut {NoStop}%
\bibitem [{\citenamefont {Chun}\ \emph {et~al.}(2015)\citenamefont {Chun},
  \citenamefont {Kim}, \citenamefont {Kim}, \citenamefont {Zheng},
  \citenamefont {Stoumpos}, \citenamefont {Malliakas}, \citenamefont
  {Mitchell}, \citenamefont {Mehlawat}, \citenamefont {Singh}, \citenamefont
  {Choi}, \citenamefont {Gog}, \citenamefont {Al-Zein}, \citenamefont {Sala},
  \citenamefont {Krisch}, \citenamefont {Chaloupka}, \citenamefont {Jackeli},
  \citenamefont {Khaliullin},\ and\ \citenamefont {Kim}}]{Chu15}%
  \BibitemOpen
  \bibfield  {author} {\bibinfo {author} {\bibfnamefont {S.~H.}\ \bibnamefont
  {Chun}}, \bibinfo {author} {\bibfnamefont {J.-W.}\ \bibnamefont {Kim}},
  \bibinfo {author} {\bibfnamefont {J.}~\bibnamefont {Kim}}, \bibinfo {author}
  {\bibfnamefont {H.}~\bibnamefont {Zheng}}, \bibinfo {author} {\bibfnamefont
  {C.}~\bibnamefont {Stoumpos}}, \bibinfo {author} {\bibfnamefont
  {C.}~\bibnamefont {Malliakas}}, \bibinfo {author} {\bibfnamefont
  {J.}~\bibnamefont {Mitchell}}, \bibinfo {author} {\bibfnamefont
  {K.}~\bibnamefont {Mehlawat}}, \bibinfo {author} {\bibfnamefont
  {Y.}~\bibnamefont {Singh}}, \bibinfo {author} {\bibfnamefont
  {Y.}~\bibnamefont {Choi}}, \bibinfo {author} {\bibfnamefont {T.}~\bibnamefont
  {Gog}}, \bibinfo {author} {\bibfnamefont {A.}~\bibnamefont {Al-Zein}},
  \bibinfo {author} {\bibfnamefont {M.}~\bibnamefont {Sala}}, \bibinfo {author}
  {\bibfnamefont {M.}~\bibnamefont {Krisch}}, \bibinfo {author} {\bibfnamefont
  {J.}~\bibnamefont {Chaloupka}}, \bibinfo {author} {\bibfnamefont
  {G.}~\bibnamefont {Jackeli}}, \bibinfo {author} {\bibfnamefont
  {G.}~\bibnamefont {Khaliullin}}, \ and\ \bibinfo {author} {\bibfnamefont
  {B.~J.}\ \bibnamefont {Kim}},\ }\href@noop {} {\bibfield  {journal} {\bibinfo
   {journal} {Nature Physics}\ }\textbf {\bibinfo {volume} {11}} (\bibinfo
  {year} {2015})}\BibitemShut {NoStop}%
\bibitem [{\citenamefont {Ruiz}\ \emph {et~al.}(2017)\citenamefont {Ruiz},
  \citenamefont {Frano}, \citenamefont {Breznay}, \citenamefont {Kimchi},
  \citenamefont {Helm}, \citenamefont {Oswald}, \citenamefont {Chan},
  \citenamefont {Birgeneau}, \citenamefont {Islam},\ and\ \citenamefont
  {Analytis}}]{Rui17}%
  \BibitemOpen
  \bibfield  {author} {\bibinfo {author} {\bibfnamefont {A.}~\bibnamefont
  {Ruiz}}, \bibinfo {author} {\bibfnamefont {A.}~\bibnamefont {Frano}},
  \bibinfo {author} {\bibfnamefont {N.}~\bibnamefont {Breznay}}, \bibinfo
  {author} {\bibfnamefont {I.}~\bibnamefont {Kimchi}}, \bibinfo {author}
  {\bibfnamefont {T.}~\bibnamefont {Helm}}, \bibinfo {author} {\bibfnamefont
  {I.}~\bibnamefont {Oswald}}, \bibinfo {author} {\bibfnamefont
  {J.}~\bibnamefont {Chan}}, \bibinfo {author} {\bibfnamefont {R.}~\bibnamefont
  {Birgeneau}}, \bibinfo {author} {\bibfnamefont {Z.}~\bibnamefont {Islam}}, \
  and\ \bibinfo {author} {\bibfnamefont {J.}~\bibnamefont {Analytis}},\
  }\href@noop {} {\bibfield  {journal} {\bibinfo  {journal} {Nat. Commun.}\
  }\textbf {\bibinfo {volume} {8}},\ \bibinfo {pages} {961} (\bibinfo {year}
  {2017})}\BibitemShut {NoStop}%
\bibitem [{\citenamefont {{Kitagawa, K. and Takayama, T. and Matsumoto, Y. and
  Kato, A. and Takano, R. and Kishimoto, Y. and Bette, S. and Dinnebier, R. and
  Jackeli, G. and Takagi, H.}}(2018)}]{Kit18}%
  \BibitemOpen
  \bibfield  {author} {\bibinfo {author} {\bibnamefont {{Kitagawa, K. and
  Takayama, T. and Matsumoto, Y. and Kato, A. and Takano, R. and Kishimoto, Y.
  and Bette, S. and Dinnebier, R. and Jackeli, G. and Takagi, H.}}},\ }\href
  {\doibase {10.1038/nature25482}} {\bibfield  {journal} {\bibinfo  {journal}
  {{Nature}}\ }\textbf {\bibinfo {volume} {{554}}},\ \bibinfo {pages} {{341}}
  (\bibinfo {year} {{2018}})}\BibitemShut {NoStop}%
\bibitem [{\citenamefont {Plumb}\ \emph {et~al.}(2014)\citenamefont {Plumb},
  \citenamefont {Clancy}, \citenamefont {Sandilands}, \citenamefont {Shankar},
  \citenamefont {Hu}, \citenamefont {Burch}, \citenamefont {Kee},\ and\
  \citenamefont {Kim}}]{Plu14}%
  \BibitemOpen
  \bibfield  {author} {\bibinfo {author} {\bibfnamefont {K.~W.}\ \bibnamefont
  {Plumb}}, \bibinfo {author} {\bibfnamefont {J.~P.}\ \bibnamefont {Clancy}},
  \bibinfo {author} {\bibfnamefont {L.~J.}\ \bibnamefont {Sandilands}},
  \bibinfo {author} {\bibfnamefont {V.~V.}\ \bibnamefont {Shankar}}, \bibinfo
  {author} {\bibfnamefont {Y.~F.}\ \bibnamefont {Hu}}, \bibinfo {author}
  {\bibfnamefont {K.~S.}\ \bibnamefont {Burch}}, \bibinfo {author}
  {\bibfnamefont {H.-Y.}\ \bibnamefont {Kee}}, \ and\ \bibinfo {author}
  {\bibfnamefont {Y.-J.}\ \bibnamefont {Kim}},\ }\href {\doibase
  10.1103/PhysRevB.90.041112} {\bibfield  {journal} {\bibinfo  {journal} {Phys.
  Rev. B}\ }\textbf {\bibinfo {volume} {90}},\ \bibinfo {pages} {041112(R)}
  (\bibinfo {year} {2014})}\BibitemShut {NoStop}%
\bibitem [{\citenamefont {Kubota}\ \emph {et~al.}(2015)\citenamefont {Kubota},
  \citenamefont {Tanaka}, \citenamefont {Ono}, \citenamefont {Narumi},\ and\
  \citenamefont {Kindo}}]{Kub15}%
  \BibitemOpen
  \bibfield  {author} {\bibinfo {author} {\bibfnamefont {Y.}~\bibnamefont
  {Kubota}}, \bibinfo {author} {\bibfnamefont {H.}~\bibnamefont {Tanaka}},
  \bibinfo {author} {\bibfnamefont {T.}~\bibnamefont {Ono}}, \bibinfo {author}
  {\bibfnamefont {Y.}~\bibnamefont {Narumi}}, \ and\ \bibinfo {author}
  {\bibfnamefont {K.}~\bibnamefont {Kindo}},\ }\href {\doibase
  10.1103/PhysRevB.91.094422} {\bibfield  {journal} {\bibinfo  {journal} {Phys.
  Rev. B}\ }\textbf {\bibinfo {volume} {91}},\ \bibinfo {pages} {094422}
  (\bibinfo {year} {2015})}\BibitemShut {NoStop}%
\bibitem [{\citenamefont {Johnson}\ \emph {et~al.}(2015)\citenamefont
  {Johnson}, \citenamefont {Williams}, \citenamefont {Haghighirad},
  \citenamefont {Singleton}, \citenamefont {Zapf}, \citenamefont {Manuel},
  \citenamefont {Mazin}, \citenamefont {Li}, \citenamefont {Jeschke},
  \citenamefont {Valent\'{\i}},\ and\ \citenamefont {Coldea}}]{Joh15}%
  \BibitemOpen
  \bibfield  {author} {\bibinfo {author} {\bibfnamefont {R.~D.}\ \bibnamefont
  {Johnson}}, \bibinfo {author} {\bibfnamefont {S.~C.}\ \bibnamefont
  {Williams}}, \bibinfo {author} {\bibfnamefont {A.~A.}\ \bibnamefont
  {Haghighirad}}, \bibinfo {author} {\bibfnamefont {J.}~\bibnamefont
  {Singleton}}, \bibinfo {author} {\bibfnamefont {V.}~\bibnamefont {Zapf}},
  \bibinfo {author} {\bibfnamefont {P.}~\bibnamefont {Manuel}}, \bibinfo
  {author} {\bibfnamefont {I.~I.}\ \bibnamefont {Mazin}}, \bibinfo {author}
  {\bibfnamefont {Y.}~\bibnamefont {Li}}, \bibinfo {author} {\bibfnamefont
  {H.~O.}\ \bibnamefont {Jeschke}}, \bibinfo {author} {\bibfnamefont
  {R.}~\bibnamefont {Valent\'{\i}}}, \ and\ \bibinfo {author} {\bibfnamefont
  {R.}~\bibnamefont {Coldea}},\ }\href {\doibase 10.1103/PhysRevB.92.235119}
  {\bibfield  {journal} {\bibinfo  {journal} {Phys. Rev. B}\ }\textbf {\bibinfo
  {volume} {92}},\ \bibinfo {pages} {235119} (\bibinfo {year}
  {2015})}\BibitemShut {NoStop}%
\bibitem [{\citenamefont {Sears}\ \emph {et~al.}(2015)\citenamefont {Sears},
  \citenamefont {Songvilay}, \citenamefont {Plumb}, \citenamefont {Clancy},
  \citenamefont {Qiu}, \citenamefont {Zhao}, \citenamefont {Parshall},\ and\
  \citenamefont {Kim}}]{Sea15}%
  \BibitemOpen
  \bibfield  {author} {\bibinfo {author} {\bibfnamefont {J.~A.}\ \bibnamefont
  {Sears}}, \bibinfo {author} {\bibfnamefont {M.}~\bibnamefont {Songvilay}},
  \bibinfo {author} {\bibfnamefont {K.~W.}\ \bibnamefont {Plumb}}, \bibinfo
  {author} {\bibfnamefont {J.~P.}\ \bibnamefont {Clancy}}, \bibinfo {author}
  {\bibfnamefont {Y.}~\bibnamefont {Qiu}}, \bibinfo {author} {\bibfnamefont
  {Y.}~\bibnamefont {Zhao}}, \bibinfo {author} {\bibfnamefont {D.}~\bibnamefont
  {Parshall}}, \ and\ \bibinfo {author} {\bibfnamefont {Y.-J.}\ \bibnamefont
  {Kim}},\ }\href {\doibase 10.1103/PhysRevB.91.144420} {\bibfield  {journal}
  {\bibinfo  {journal} {Phys. Rev. B}\ }\textbf {\bibinfo {volume} {91}},\
  \bibinfo {pages} {144420} (\bibinfo {year} {2015})}\BibitemShut {NoStop}%
\bibitem [{\citenamefont {Cao}\ \emph {et~al.}(2016)\citenamefont {Cao},
  \citenamefont {Banerjee}, \citenamefont {Yan}, \citenamefont {Bridges},
  \citenamefont {Lumsden}, \citenamefont {Mandrus}, \citenamefont {Tennant},
  \citenamefont {Chakoumakos},\ and\ \citenamefont {Nagler}}]{Cao16}%
  \BibitemOpen
  \bibfield  {author} {\bibinfo {author} {\bibfnamefont {H.~B.}\ \bibnamefont
  {Cao}}, \bibinfo {author} {\bibfnamefont {A.}~\bibnamefont {Banerjee}},
  \bibinfo {author} {\bibfnamefont {J.-Q.}\ \bibnamefont {Yan}}, \bibinfo
  {author} {\bibfnamefont {C.~A.}\ \bibnamefont {Bridges}}, \bibinfo {author}
  {\bibfnamefont {M.~D.}\ \bibnamefont {Lumsden}}, \bibinfo {author}
  {\bibfnamefont {D.~G.}\ \bibnamefont {Mandrus}}, \bibinfo {author}
  {\bibfnamefont {D.~A.}\ \bibnamefont {Tennant}}, \bibinfo {author}
  {\bibfnamefont {B.~C.}\ \bibnamefont {Chakoumakos}}, \ and\ \bibinfo {author}
  {\bibfnamefont {S.~E.}\ \bibnamefont {Nagler}},\ }\href {\doibase
  10.1103/PhysRevB.93.134423} {\bibfield  {journal} {\bibinfo  {journal} {Phys.
  Rev. B}\ }\textbf {\bibinfo {volume} {93}},\ \bibinfo {pages} {134423}
  (\bibinfo {year} {2016})}\BibitemShut {NoStop}%
\bibitem [{\citenamefont {Park}\ \emph {et~al.}(2016)\citenamefont {Park},
  \citenamefont {Do}, \citenamefont {Choi}, \citenamefont {nd~T.-H.~Jang},
  \citenamefont {Schefer}, \citenamefont {Wu}, \citenamefont {Gardner},
  \citenamefont {Park}, \citenamefont {Park},\ and\ \citenamefont
  {Ji}}]{Par16}%
  \BibitemOpen
  \bibfield  {author} {\bibinfo {author} {\bibfnamefont {S.-Y.}\ \bibnamefont
  {Park}}, \bibinfo {author} {\bibfnamefont {S.-H.}\ \bibnamefont {Do}},
  \bibinfo {author} {\bibfnamefont {K.-Y.}\ \bibnamefont {Choi}}, \bibinfo
  {author} {\bibfnamefont {D.~J.}\ \bibnamefont {nd~T.-H.~Jang}}, \bibinfo
  {author} {\bibfnamefont {J.}~\bibnamefont {Schefer}}, \bibinfo {author}
  {\bibfnamefont {C.-M.}\ \bibnamefont {Wu}}, \bibinfo {author} {\bibfnamefont
  {J.~S.}\ \bibnamefont {Gardner}}, \bibinfo {author} {\bibfnamefont
  {J.~M.~S.}\ \bibnamefont {Park}}, \bibinfo {author} {\bibfnamefont {J.-H.}\
  \bibnamefont {Park}}, \ and\ \bibinfo {author} {\bibfnamefont
  {S.}~\bibnamefont {Ji}},\ }\href {http://arxiv.org/abs/1609.05690} {\bibfield
   {journal} {\bibinfo  {journal} {arXiv:1609.05690}\ } (\bibinfo {year}
  {2016})}\BibitemShut {NoStop}%
\bibitem [{\citenamefont {Yadav}\ \emph {et~al.}(2016)\citenamefont {Yadav},
  \citenamefont {Bogdanov}, \citenamefont {Katukuri}, \citenamefont
  {Nishimoto}, \citenamefont {van~den Brink},\ and\ \citenamefont
  {Hozoi}}]{Yad16}%
  \BibitemOpen
  \bibfield  {author} {\bibinfo {author} {\bibfnamefont {R.}~\bibnamefont
  {Yadav}}, \bibinfo {author} {\bibfnamefont {N.~A.}\ \bibnamefont {Bogdanov}},
  \bibinfo {author} {\bibfnamefont {V.~M.}\ \bibnamefont {Katukuri}}, \bibinfo
  {author} {\bibfnamefont {S.}~\bibnamefont {Nishimoto}}, \bibinfo {author}
  {\bibfnamefont {J.}~\bibnamefont {van~den Brink}}, \ and\ \bibinfo {author}
  {\bibfnamefont {L.}~\bibnamefont {Hozoi}},\ }\href {\doibase
  {10.1038/srep37925}} {\bibfield  {journal} {\bibinfo  {journal} {Scientific
  Reports}\ }\textbf {\bibinfo {volume} {{6}}},\ \bibinfo {pages} {{37925}}
  (\bibinfo {year} {{2016}})}\BibitemShut {NoStop}%
\bibitem [{\citenamefont {Sandilands}\ \emph {et~al.}(2015)\citenamefont
  {Sandilands}, \citenamefont {Tian}, \citenamefont {Plumb}, \citenamefont
  {Kim},\ and\ \citenamefont {Burch}}]{San15}%
  \BibitemOpen
  \bibfield  {author} {\bibinfo {author} {\bibfnamefont {L.~J.}\ \bibnamefont
  {Sandilands}}, \bibinfo {author} {\bibfnamefont {Y.}~\bibnamefont {Tian}},
  \bibinfo {author} {\bibfnamefont {K.~W.}\ \bibnamefont {Plumb}}, \bibinfo
  {author} {\bibfnamefont {Y.-J.}\ \bibnamefont {Kim}}, \ and\ \bibinfo
  {author} {\bibfnamefont {K.~S.}\ \bibnamefont {Burch}},\ }\href {\doibase
  10.1103/PhysRevLett.114.147201} {\bibfield  {journal} {\bibinfo  {journal}
  {Phys. Rev. Lett.}\ }\textbf {\bibinfo {volume} {114}},\ \bibinfo {pages}
  {147201} (\bibinfo {year} {2015})}\BibitemShut {NoStop}%
\bibitem [{\citenamefont {Ponomaryov}\ \emph {et~al.}(2017)\citenamefont
  {Ponomaryov}, \citenamefont {Schulze}, \citenamefont {Wosnitza},
  \citenamefont {Lampen-Kelley}, \citenamefont {Banerjee}, \citenamefont {Yan},
  \citenamefont {Bridges}, \citenamefont {Mandrus}, \citenamefont {Nagler},
  \citenamefont {Kolezhuk},\ and\ \citenamefont {Zvyagin}}]{Pon17}%
  \BibitemOpen
  \bibfield  {author} {\bibinfo {author} {\bibfnamefont {A.~N.}\ \bibnamefont
  {Ponomaryov}}, \bibinfo {author} {\bibfnamefont {E.}~\bibnamefont {Schulze}},
  \bibinfo {author} {\bibfnamefont {J.}~\bibnamefont {Wosnitza}}, \bibinfo
  {author} {\bibfnamefont {P.}~\bibnamefont {Lampen-Kelley}}, \bibinfo {author}
  {\bibfnamefont {A.}~\bibnamefont {Banerjee}}, \bibinfo {author}
  {\bibfnamefont {J.-Q.}\ \bibnamefont {Yan}}, \bibinfo {author} {\bibfnamefont
  {C.~A.}\ \bibnamefont {Bridges}}, \bibinfo {author} {\bibfnamefont {D.~G.}\
  \bibnamefont {Mandrus}}, \bibinfo {author} {\bibfnamefont {S.~E.}\
  \bibnamefont {Nagler}}, \bibinfo {author} {\bibfnamefont {A.~K.}\
  \bibnamefont {Kolezhuk}}, \ and\ \bibinfo {author} {\bibfnamefont {S.~A.}\
  \bibnamefont {Zvyagin}},\ }\href {\doibase 10.1103/PhysRevB.96.241107}
  {\bibfield  {journal} {\bibinfo  {journal} {Phys. Rev. B}\ }\textbf {\bibinfo
  {volume} {96}},\ \bibinfo {pages} {241107(R)} (\bibinfo {year}
  {2017})}\BibitemShut {NoStop}%
\bibitem [{\citenamefont {Wang}\ \emph {et~al.}(2017)\citenamefont {Wang},
  \citenamefont {Reschke}, \citenamefont {H\"uvonen}, \citenamefont {Do},
  \citenamefont {Choi}, \citenamefont {Gensch}, \citenamefont {Nagel},
  \citenamefont {Room},\ and\ \citenamefont {Loidl}}]{Wan17}%
  \BibitemOpen
  \bibfield  {author} {\bibinfo {author} {\bibfnamefont {Z.}~\bibnamefont
  {Wang}}, \bibinfo {author} {\bibfnamefont {S.}~\bibnamefont {Reschke}},
  \bibinfo {author} {\bibfnamefont {D.}~\bibnamefont {H\"uvonen}}, \bibinfo
  {author} {\bibfnamefont {S.-H.}\ \bibnamefont {Do}}, \bibinfo {author}
  {\bibfnamefont {K.-Y.}\ \bibnamefont {Choi}}, \bibinfo {author}
  {\bibfnamefont {M.}~\bibnamefont {Gensch}}, \bibinfo {author} {\bibfnamefont
  {U.}~\bibnamefont {Nagel}}, \bibinfo {author} {\bibfnamefont
  {T.}~\bibnamefont {Room}}, \ and\ \bibinfo {author} {\bibfnamefont
  {A.}~\bibnamefont {Loidl}},\ }\href {\doibase 10.1103/PhysRevLett.119.227202}
  {\bibfield  {journal} {\bibinfo  {journal} {Phys. Rev. Lett.}\ }\textbf
  {\bibinfo {volume} {119}},\ \bibinfo {pages} {227202} (\bibinfo {year}
  {2017})}\BibitemShut {NoStop}%
\bibitem [{\citenamefont {Little}\ \emph {et~al.}(2017)\citenamefont {Little},
  \citenamefont {Wu}, \citenamefont {Lampen-Kelley}, \citenamefont {Banerjee},
  \citenamefont {Patankar}, \citenamefont {Rees}, \citenamefont {Bridges},
  \citenamefont {Yan}, \citenamefont {Mandrus}, \citenamefont {Nagler},\ and\
  \citenamefont {Orenstein}}]{Lit17}%
  \BibitemOpen
  \bibfield  {author} {\bibinfo {author} {\bibfnamefont {A.}~\bibnamefont
  {Little}}, \bibinfo {author} {\bibfnamefont {L.}~\bibnamefont {Wu}}, \bibinfo
  {author} {\bibfnamefont {P.}~\bibnamefont {Lampen-Kelley}}, \bibinfo {author}
  {\bibfnamefont {A.}~\bibnamefont {Banerjee}}, \bibinfo {author}
  {\bibfnamefont {S.}~\bibnamefont {Patankar}}, \bibinfo {author}
  {\bibfnamefont {D.}~\bibnamefont {Rees}}, \bibinfo {author} {\bibfnamefont
  {C.~A.}\ \bibnamefont {Bridges}}, \bibinfo {author} {\bibfnamefont {J.-Q.}\
  \bibnamefont {Yan}}, \bibinfo {author} {\bibfnamefont {D.}~\bibnamefont
  {Mandrus}}, \bibinfo {author} {\bibfnamefont {S.~E.}\ \bibnamefont {Nagler}},
  \ and\ \bibinfo {author} {\bibfnamefont {J.}~\bibnamefont {Orenstein}},\
  }\href {\doibase 10.1103/PhysRevLett.119.227201} {\bibfield  {journal}
  {\bibinfo  {journal} {Phys. Rev. Lett.}\ }\textbf {\bibinfo {volume} {119}},\
  \bibinfo {pages} {227201} (\bibinfo {year} {2017})}\BibitemShut {NoStop}%
\bibitem [{\citenamefont {Wu}\ \emph {et~al.}(2018)\citenamefont {Wu},
  \citenamefont {Little}, \citenamefont {Aldape}, \citenamefont {Rees},
  \citenamefont {Thewalt}, \citenamefont {Lampen-Kelley}, \citenamefont
  {Banerjee}, \citenamefont {Bridges}, \citenamefont {Yan}, \citenamefont
  {Boone}, \citenamefont {Patankar}, \citenamefont {Goldhaber-Gordon},
  \citenamefont {Mandrus}, \citenamefont {Nagler}, \citenamefont {Altman},\
  and\ \citenamefont {Orenstein}}]{Wu18}%
  \BibitemOpen
  \bibfield  {author} {\bibinfo {author} {\bibfnamefont {L.}~\bibnamefont
  {Wu}}, \bibinfo {author} {\bibfnamefont {A.}~\bibnamefont {Little}}, \bibinfo
  {author} {\bibfnamefont {E.~E.}\ \bibnamefont {Aldape}}, \bibinfo {author}
  {\bibfnamefont {D.}~\bibnamefont {Rees}}, \bibinfo {author} {\bibfnamefont
  {E.}~\bibnamefont {Thewalt}}, \bibinfo {author} {\bibfnamefont
  {P.}~\bibnamefont {Lampen-Kelley}}, \bibinfo {author} {\bibfnamefont
  {A.}~\bibnamefont {Banerjee}}, \bibinfo {author} {\bibfnamefont {C.~A.}\
  \bibnamefont {Bridges}}, \bibinfo {author} {\bibfnamefont {J.-Q.}\
  \bibnamefont {Yan}}, \bibinfo {author} {\bibfnamefont {D.}~\bibnamefont
  {Boone}}, \bibinfo {author} {\bibfnamefont {S.}~\bibnamefont {Patankar}},
  \bibinfo {author} {\bibfnamefont {D.}~\bibnamefont {Goldhaber-Gordon}},
  \bibinfo {author} {\bibfnamefont {D.}~\bibnamefont {Mandrus}}, \bibinfo
  {author} {\bibfnamefont {S.~E.}\ \bibnamefont {Nagler}}, \bibinfo {author}
  {\bibfnamefont {E.}~\bibnamefont {Altman}}, \ and\ \bibinfo {author}
  {\bibfnamefont {J.}~\bibnamefont {Orenstein}},\ }\href {\doibase
  10.1103/PhysRevB.98.094425} {\bibfield  {journal} {\bibinfo  {journal} {Phys.
  Rev. B}\ }\textbf {\bibinfo {volume} {98}},\ \bibinfo {pages} {094425}
  (\bibinfo {year} {2018})}\BibitemShut {NoStop}%
\bibitem [{\citenamefont {Shi}\ \emph {et~al.}(2018)\citenamefont {Shi},
  \citenamefont {Liu}, \citenamefont {Lin}, \citenamefont {Zhang},
  \citenamefont {Zhang}, \citenamefont {Wang}, \citenamefont {Shi},
  \citenamefont {Dong},\ and\ \citenamefont {Wang}}]{Shi18}%
  \BibitemOpen
  \bibfield  {author} {\bibinfo {author} {\bibfnamefont {L.~Y.}\ \bibnamefont
  {Shi}}, \bibinfo {author} {\bibfnamefont {Y.~Q.}\ \bibnamefont {Liu}},
  \bibinfo {author} {\bibfnamefont {T.}~\bibnamefont {Lin}}, \bibinfo {author}
  {\bibfnamefont {M.~Y.}\ \bibnamefont {Zhang}}, \bibinfo {author}
  {\bibfnamefont {S.~J.}\ \bibnamefont {Zhang}}, \bibinfo {author}
  {\bibfnamefont {L.}~\bibnamefont {Wang}}, \bibinfo {author} {\bibfnamefont
  {Y.~G.}\ \bibnamefont {Shi}}, \bibinfo {author} {\bibfnamefont
  {T.}~\bibnamefont {Dong}}, \ and\ \bibinfo {author} {\bibfnamefont {N.~L.}\
  \bibnamefont {Wang}},\ }\href {\doibase 10.1103/PhysRevB.98.094414}
  {\bibfield  {journal} {\bibinfo  {journal} {Phys. Rev. B}\ }\textbf {\bibinfo
  {volume} {98}},\ \bibinfo {pages} {094414} (\bibinfo {year}
  {2018})}\BibitemShut {NoStop}%
\bibitem [{\citenamefont {Wellm}\ \emph {et~al.}(2018)\citenamefont {Wellm},
  \citenamefont {Zeisner}, \citenamefont {Alfonsov}, \citenamefont {Wolter},
  \citenamefont {Roslova}, \citenamefont {Isaeva}, \citenamefont {Doert},
  \citenamefont {Vojta}, \citenamefont {B\"uchner},\ and\ \citenamefont
  {Kataev}}]{Wel18}%
  \BibitemOpen
  \bibfield  {author} {\bibinfo {author} {\bibfnamefont {C.}~\bibnamefont
  {Wellm}}, \bibinfo {author} {\bibfnamefont {J.}~\bibnamefont {Zeisner}},
  \bibinfo {author} {\bibfnamefont {A.}~\bibnamefont {Alfonsov}}, \bibinfo
  {author} {\bibfnamefont {A.~U.~B.}\ \bibnamefont {Wolter}}, \bibinfo {author}
  {\bibfnamefont {M.}~\bibnamefont {Roslova}}, \bibinfo {author} {\bibfnamefont
  {A.}~\bibnamefont {Isaeva}}, \bibinfo {author} {\bibfnamefont
  {T.}~\bibnamefont {Doert}}, \bibinfo {author} {\bibfnamefont
  {M.}~\bibnamefont {Vojta}}, \bibinfo {author} {\bibfnamefont
  {B.}~\bibnamefont {B\"uchner}}, \ and\ \bibinfo {author} {\bibfnamefont
  {V.}~\bibnamefont {Kataev}},\ }\href {\doibase 10.1103/PhysRevB.98.184408}
  {\bibfield  {journal} {\bibinfo  {journal} {Phys. Rev. B}\ }\textbf {\bibinfo
  {volume} {98}},\ \bibinfo {pages} {184408} (\bibinfo {year}
  {2018})}\BibitemShut {NoStop}%
\bibitem [{\citenamefont {Banerjee}\ \emph {et~al.}(2016)\citenamefont
  {Banerjee}, \citenamefont {Bridges}, \citenamefont {Yan}, \citenamefont
  {Aczel}, \citenamefont {Li}, \citenamefont {Stone}, \citenamefont {Granroth},
  \citenamefont {Lumsden}, \citenamefont {Yiu}, \citenamefont {Knolle},
  \citenamefont {Bhattacharjee}, \citenamefont {Kovrizhin}, \citenamefont
  {Moessner}, \citenamefont {Tennant}, \citenamefont {Mandrus},\ and\
  \citenamefont {Nagler}}]{Ban16}%
  \BibitemOpen
  \bibfield  {author} {\bibinfo {author} {\bibfnamefont {A.}~\bibnamefont
  {Banerjee}}, \bibinfo {author} {\bibfnamefont {C.~A.}\ \bibnamefont
  {Bridges}}, \bibinfo {author} {\bibfnamefont {J.-Q.}\ \bibnamefont {Yan}},
  \bibinfo {author} {\bibfnamefont {A.~A.}\ \bibnamefont {Aczel}}, \bibinfo
  {author} {\bibfnamefont {L.}~\bibnamefont {Li}}, \bibinfo {author}
  {\bibfnamefont {M.~B.}\ \bibnamefont {Stone}}, \bibinfo {author}
  {\bibfnamefont {G.~E.}\ \bibnamefont {Granroth}}, \bibinfo {author}
  {\bibfnamefont {M.~D.}\ \bibnamefont {Lumsden}}, \bibinfo {author}
  {\bibfnamefont {Y.}~\bibnamefont {Yiu}}, \bibinfo {author} {\bibfnamefont
  {J.}~\bibnamefont {Knolle}}, \bibinfo {author} {\bibfnamefont
  {S.}~\bibnamefont {Bhattacharjee}}, \bibinfo {author} {\bibfnamefont {D.~L.}\
  \bibnamefont {Kovrizhin}}, \bibinfo {author} {\bibfnamefont {R.}~\bibnamefont
  {Moessner}}, \bibinfo {author} {\bibfnamefont {D.~A.}\ \bibnamefont
  {Tennant}}, \bibinfo {author} {\bibfnamefont {D.~G.}\ \bibnamefont
  {Mandrus}}, \ and\ \bibinfo {author} {\bibfnamefont {S.~E.}\ \bibnamefont
  {Nagler}},\ }\href {\doibase 10.1038/nmat4604} {\bibfield  {journal}
  {\bibinfo  {journal} {Nat. Mater.}\ }\textbf {\bibinfo {volume} {15}},\
  \bibinfo {pages} {733} (\bibinfo {year} {2016})}\BibitemShut {NoStop}%
\bibitem [{\citenamefont {Banerjee}\ \emph {et~al.}(2017)\citenamefont
  {Banerjee}, \citenamefont {Yan}, \citenamefont {Knolle}, \citenamefont
  {Bridges}, \citenamefont {Stone}, \citenamefont {Lumsden}, \citenamefont
  {Mandrus}, \citenamefont {Tennant}, \citenamefont {Moessner},\ and\
  \citenamefont {Nagler}}]{Ban17}%
  \BibitemOpen
  \bibfield  {author} {\bibinfo {author} {\bibfnamefont {A.}~\bibnamefont
  {Banerjee}}, \bibinfo {author} {\bibfnamefont {J.}~\bibnamefont {Yan}},
  \bibinfo {author} {\bibfnamefont {J.}~\bibnamefont {Knolle}}, \bibinfo
  {author} {\bibfnamefont {C.~A.}\ \bibnamefont {Bridges}}, \bibinfo {author}
  {\bibfnamefont {M.~B.}\ \bibnamefont {Stone}}, \bibinfo {author}
  {\bibfnamefont {M.~D.}\ \bibnamefont {Lumsden}}, \bibinfo {author}
  {\bibfnamefont {D.~G.}\ \bibnamefont {Mandrus}}, \bibinfo {author}
  {\bibfnamefont {D.~A.}\ \bibnamefont {Tennant}}, \bibinfo {author}
  {\bibfnamefont {R.}~\bibnamefont {Moessner}}, \ and\ \bibinfo {author}
  {\bibfnamefont {S.~E.}\ \bibnamefont {Nagler}},\ }\href {\doibase
  10.1126/science.aah6015} {\bibfield  {journal} {\bibinfo  {journal}
  {Science}\ }\textbf {\bibinfo {volume} {356}},\ \bibinfo {pages} {1055}
  (\bibinfo {year} {2017})}\BibitemShut {NoStop}%
\bibitem [{\citenamefont {Ran}\ \emph {et~al.}(2017)\citenamefont {Ran},
  \citenamefont {Wang}, \citenamefont {Wang}, \citenamefont {Dong},
  \citenamefont {Ren}, \citenamefont {Bao}, \citenamefont {Li}, \citenamefont
  {Ma}, \citenamefont {Gan}, \citenamefont {Zhang}, \citenamefont {Park},
  \citenamefont {Deng}, \citenamefont {Danilkin}, \citenamefont {Yu},
  \citenamefont {Li},\ and\ \citenamefont {Wen}}]{Ran17}%
  \BibitemOpen
  \bibfield  {author} {\bibinfo {author} {\bibfnamefont {K.}~\bibnamefont
  {Ran}}, \bibinfo {author} {\bibfnamefont {J.}~\bibnamefont {Wang}}, \bibinfo
  {author} {\bibfnamefont {W.}~\bibnamefont {Wang}}, \bibinfo {author}
  {\bibfnamefont {Z.-Y.}\ \bibnamefont {Dong}}, \bibinfo {author}
  {\bibfnamefont {X.}~\bibnamefont {Ren}}, \bibinfo {author} {\bibfnamefont
  {S.}~\bibnamefont {Bao}}, \bibinfo {author} {\bibfnamefont {S.}~\bibnamefont
  {Li}}, \bibinfo {author} {\bibfnamefont {Z.}~\bibnamefont {Ma}}, \bibinfo
  {author} {\bibfnamefont {Y.}~\bibnamefont {Gan}}, \bibinfo {author}
  {\bibfnamefont {Y.}~\bibnamefont {Zhang}}, \bibinfo {author} {\bibfnamefont
  {J.~T.}\ \bibnamefont {Park}}, \bibinfo {author} {\bibfnamefont
  {G.}~\bibnamefont {Deng}}, \bibinfo {author} {\bibfnamefont {S.}~\bibnamefont
  {Danilkin}}, \bibinfo {author} {\bibfnamefont {S.-L.}\ \bibnamefont {Yu}},
  \bibinfo {author} {\bibfnamefont {J.-X.}\ \bibnamefont {Li}}, \ and\ \bibinfo
  {author} {\bibfnamefont {J.}~\bibnamefont {Wen}},\ }\href {\doibase
  10.1103/PhysRevLett.118.107203} {\bibfield  {journal} {\bibinfo  {journal}
  {Phys. Rev. Lett.}\ }\textbf {\bibinfo {volume} {118}},\ \bibinfo {pages}
  {107203} (\bibinfo {year} {2017})}\BibitemShut {NoStop}%
\bibitem [{\citenamefont {Do}\ \emph {et~al.}(2017)\citenamefont {Do},
  \citenamefont {Park}, \citenamefont {Yoshitake}, \citenamefont {Nasu},
  \citenamefont {Motome}, \citenamefont {Kwon}, \citenamefont {Adroja},
  \citenamefont {Voneshen}, \citenamefont {Kim}, \citenamefont {Jang},
  \citenamefont {Park}, \citenamefont {Choi},\ and\ \citenamefont {Ji}}]{Do17}%
  \BibitemOpen
  \bibfield  {author} {\bibinfo {author} {\bibfnamefont {S.-H.}\ \bibnamefont
  {Do}}, \bibinfo {author} {\bibfnamefont {S.-Y.}\ \bibnamefont {Park}},
  \bibinfo {author} {\bibfnamefont {J.}~\bibnamefont {Yoshitake}}, \bibinfo
  {author} {\bibfnamefont {J.}~\bibnamefont {Nasu}}, \bibinfo {author}
  {\bibfnamefont {Y.}~\bibnamefont {Motome}}, \bibinfo {author} {\bibfnamefont
  {Y.~S.}\ \bibnamefont {Kwon}}, \bibinfo {author} {\bibfnamefont {D.~T.}\
  \bibnamefont {Adroja}}, \bibinfo {author} {\bibfnamefont {D.~J.}\
  \bibnamefont {Voneshen}}, \bibinfo {author} {\bibfnamefont {K.}~\bibnamefont
  {Kim}}, \bibinfo {author} {\bibfnamefont {T.~H.}\ \bibnamefont {Jang}},
  \bibinfo {author} {\bibfnamefont {J.~H.}\ \bibnamefont {Park}}, \bibinfo
  {author} {\bibfnamefont {K.-Y.}\ \bibnamefont {Choi}}, \ and\ \bibinfo
  {author} {\bibfnamefont {S.}~\bibnamefont {Ji}},\ }\href {\doibase
  {10.1038/nphys4264}} {\bibfield  {journal} {\bibinfo  {journal} {{Nature
  Physics}}\ }\textbf {\bibinfo {volume} {{13}}},\ \bibinfo {pages} {{1079}}
  (\bibinfo {year} {{2017}})}\BibitemShut {NoStop}%
\bibitem [{\citenamefont {Banerjee}\ \emph {et~al.}(2018)\citenamefont
  {Banerjee}, \citenamefont {Lampen-Kelley}, \citenamefont {Knolle},
  \citenamefont {Balz}, \citenamefont {Aczel}, \citenamefont {Winn},
  \citenamefont {Liu}, \citenamefont {Pajerowski}, \citenamefont {Yan},
  \citenamefont {Bridges}, \citenamefont {Savici}, \citenamefont {Chakoumakos},
  \citenamefont {Lumsden}, \citenamefont {Tennant}, \citenamefont {Moessner},
  \citenamefont {Mandrus},\ and\ \citenamefont {Nagler}}]{Ban18}%
  \BibitemOpen
  \bibfield  {author} {\bibinfo {author} {\bibfnamefont {A.}~\bibnamefont
  {Banerjee}}, \bibinfo {author} {\bibfnamefont {P.}~\bibnamefont
  {Lampen-Kelley}}, \bibinfo {author} {\bibfnamefont {J.}~\bibnamefont
  {Knolle}}, \bibinfo {author} {\bibfnamefont {C.}~\bibnamefont {Balz}},
  \bibinfo {author} {\bibfnamefont {A.~A.}\ \bibnamefont {Aczel}}, \bibinfo
  {author} {\bibfnamefont {B.}~\bibnamefont {Winn}}, \bibinfo {author}
  {\bibfnamefont {Y.}~\bibnamefont {Liu}}, \bibinfo {author} {\bibfnamefont
  {D.}~\bibnamefont {Pajerowski}}, \bibinfo {author} {\bibfnamefont
  {J.}~\bibnamefont {Yan}}, \bibinfo {author} {\bibfnamefont {C.~A.}\
  \bibnamefont {Bridges}}, \bibinfo {author} {\bibfnamefont {A.~T.}\
  \bibnamefont {Savici}}, \bibinfo {author} {\bibfnamefont {B.~C.}\
  \bibnamefont {Chakoumakos}}, \bibinfo {author} {\bibfnamefont {M.~D.}\
  \bibnamefont {Lumsden}}, \bibinfo {author} {\bibfnamefont {D.~A.}\
  \bibnamefont {Tennant}}, \bibinfo {author} {\bibfnamefont {R.}~\bibnamefont
  {Moessner}}, \bibinfo {author} {\bibfnamefont {D.~G.}\ \bibnamefont
  {Mandrus}}, \ and\ \bibinfo {author} {\bibfnamefont {S.~E.}\ \bibnamefont
  {Nagler}},\ }\href {\doibase 10.1038/s41535-018-0079-2} {\bibfield  {journal}
  {\bibinfo  {journal} {npj Quantum Materials}\ }\textbf {\bibinfo {volume}
  {3}},\ \bibinfo {pages} {8} (\bibinfo {year} {2018})}\BibitemShut {NoStop}%
\bibitem [{\citenamefont {Kasahara}\ \emph {et~al.}(2018)\citenamefont
  {Kasahara}, \citenamefont {Ohnishi}, \citenamefont {Mizukami}, \citenamefont
  {Tanaka}, \citenamefont {Ma}, \citenamefont {Sugii}, \citenamefont {Kurita},
  \citenamefont {Tanaka}, \citenamefont {Nasu}, \citenamefont {Motome},
  \citenamefont {Shibauchi},\ and\ \citenamefont {Matsuda}}]{Kas18}%
  \BibitemOpen
  \bibfield  {author} {\bibinfo {author} {\bibfnamefont {Y.}~\bibnamefont
  {Kasahara}}, \bibinfo {author} {\bibfnamefont {T.}~\bibnamefont {Ohnishi}},
  \bibinfo {author} {\bibfnamefont {Y.}~\bibnamefont {Mizukami}}, \bibinfo
  {author} {\bibfnamefont {O.}~\bibnamefont {Tanaka}}, \bibinfo {author}
  {\bibfnamefont {S.}~\bibnamefont {Ma}}, \bibinfo {author} {\bibfnamefont
  {K.}~\bibnamefont {Sugii}}, \bibinfo {author} {\bibfnamefont
  {N.}~\bibnamefont {Kurita}}, \bibinfo {author} {\bibfnamefont
  {H.}~\bibnamefont {Tanaka}}, \bibinfo {author} {\bibfnamefont
  {J.}~\bibnamefont {Nasu}}, \bibinfo {author} {\bibfnamefont {Y.}~\bibnamefont
  {Motome}}, \bibinfo {author} {\bibfnamefont {T.}~\bibnamefont {Shibauchi}}, \
  and\ \bibinfo {author} {\bibfnamefont {Y.}~\bibnamefont {Matsuda}},\ }\href
  {\doibase {10.1038/s41586-018-0274-0}} {\bibfield  {journal} {\bibinfo
  {journal} {{Nature}}\ }\textbf {\bibinfo {volume} {{559}}},\ \bibinfo {pages}
  {{227}} (\bibinfo {year} {{2018}})}\BibitemShut {NoStop}%
\bibitem [{\citenamefont {Ye}\ \emph {et~al.}(2018)\citenamefont {Ye},
  \citenamefont {Halasz}, \citenamefont {Savary},\ and\ \citenamefont
  {Balents}}]{Ye18}%
  \BibitemOpen
  \bibfield  {author} {\bibinfo {author} {\bibfnamefont {M.}~\bibnamefont
  {Ye}}, \bibinfo {author} {\bibfnamefont {G.~B.}\ \bibnamefont {Halasz}},
  \bibinfo {author} {\bibfnamefont {L.}~\bibnamefont {Savary}}, \ and\ \bibinfo
  {author} {\bibfnamefont {L.}~\bibnamefont {Balents}},\ }\href {\doibase
  10.1103/PhysRevLett.121.147201} {\bibfield  {journal} {\bibinfo  {journal}
  {Phys. Rev. Lett.}\ }\textbf {\bibinfo {volume} {121}},\ \bibinfo {pages}
  {147201} (\bibinfo {year} {2018})}\BibitemShut {NoStop}%
\bibitem [{\citenamefont {Cookmeyer}\ and\ \citenamefont
  {Moore}(2018)}]{Coo18}%
  \BibitemOpen
  \bibfield  {author} {\bibinfo {author} {\bibfnamefont {J.}~\bibnamefont
  {Cookmeyer}}\ and\ \bibinfo {author} {\bibfnamefont {J.~E.}\ \bibnamefont
  {Moore}},\ }\href {\doibase 10.1103/PhysRevB.98.060412} {\bibfield  {journal}
  {\bibinfo  {journal} {Phys. Rev. B}\ }\textbf {\bibinfo {volume} {98}},\
  \bibinfo {pages} {060412(R)} (\bibinfo {year} {2018})}\BibitemShut {NoStop}%
\bibitem [{\citenamefont {Vinkler-Aviv}\ and\ \citenamefont
  {Rosch}(2018)}]{Vin18}%
  \BibitemOpen
  \bibfield  {author} {\bibinfo {author} {\bibfnamefont {Y.}~\bibnamefont
  {Vinkler-Aviv}}\ and\ \bibinfo {author} {\bibfnamefont {A.}~\bibnamefont
  {Rosch}},\ }\href {\doibase 10.1103/PhysRevX.8.031032} {\bibfield  {journal}
  {\bibinfo  {journal} {Phys. Rev. X}\ }\textbf {\bibinfo {volume} {8}},\
  \bibinfo {pages} {031032} (\bibinfo {year} {2018})}\BibitemShut {NoStop}%
\bibitem [{\citenamefont {Jiang}\ \emph {et~al.}(2019)\citenamefont {Jiang},
  \citenamefont {Devereaux},\ and\ \citenamefont {Jiang}}]{Jia19}%
  \BibitemOpen
  \bibfield  {author} {\bibinfo {author} {\bibfnamefont {Y.-F.}\ \bibnamefont
  {Jiang}}, \bibinfo {author} {\bibfnamefont {T.~P.}\ \bibnamefont
  {Devereaux}}, \ and\ \bibinfo {author} {\bibfnamefont {H.-C.}\ \bibnamefont
  {Jiang}},\ }\href {https://arxiv.org/abs/1901.09131} {\bibfield  {journal}
  {\bibinfo  {journal} {arXiv:1901.09131}\ } (\bibinfo {year}
  {2019})}\BibitemShut {NoStop}%
\bibitem [{\citenamefont {Gordon}\ \emph {et~al.}(2019)\citenamefont {Gordon},
  \citenamefont {Catuneanu}, \citenamefont {S{\o}rensen},\ and\ \citenamefont
  {Kee}}]{Gor19}%
  \BibitemOpen
  \bibfield  {author} {\bibinfo {author} {\bibfnamefont {J.~S.}\ \bibnamefont
  {Gordon}}, \bibinfo {author} {\bibfnamefont {A.}~\bibnamefont {Catuneanu}},
  \bibinfo {author} {\bibfnamefont {E.~S.}\ \bibnamefont {S{\o}rensen}}, \ and\
  \bibinfo {author} {\bibfnamefont {H.-Y.}\ \bibnamefont {Kee}},\ }\href
  {https://arxiv.org/abs/1901.09943} {\bibfield  {journal} {\bibinfo  {journal}
  {arXiv:1901.09943}\ } (\bibinfo {year} {2019})}\BibitemShut {NoStop}%
\bibitem [{\citenamefont {Takikawa}\ and\ \citenamefont
  {Fujimoto}(2019)}]{Tak19}%
  \BibitemOpen
  \bibfield  {author} {\bibinfo {author} {\bibfnamefont {D.}~\bibnamefont
  {Takikawa}}\ and\ \bibinfo {author} {\bibfnamefont {S.}~\bibnamefont
  {Fujimoto}},\ }\href {https://arxiv.org/abs/1902.06433} {\bibfield  {journal}
  {\bibinfo  {journal} {arXiv:1902.06433}\ } (\bibinfo {year}
  {2019})}\BibitemShut {NoStop}%
\bibitem [{\citenamefont {Habicht}\ \emph {et~al.}(2015)\citenamefont
  {Habicht}, \citenamefont {Quintero-Castro}, \citenamefont {Toft-Petersen},
  \citenamefont {Kure}, \citenamefont {M\"ade}, \citenamefont {Groitl},\ and\
  \citenamefont {Le}}]{Hab15}%
  \BibitemOpen
  \bibfield  {author} {\bibinfo {author} {\bibfnamefont {K.}~\bibnamefont
  {Habicht}}, \bibinfo {author} {\bibfnamefont {D.~L.}\ \bibnamefont
  {Quintero-Castro}}, \bibinfo {author} {\bibfnamefont {R.}~\bibnamefont
  {Toft-Petersen}}, \bibinfo {author} {\bibfnamefont {M.}~\bibnamefont {Kure}},
  \bibinfo {author} {\bibfnamefont {L.}~\bibnamefont {M\"ade}}, \bibinfo
  {author} {\bibfnamefont {F.}~\bibnamefont {Groitl}}, \ and\ \bibinfo {author}
  {\bibfnamefont {M.~D.}\ \bibnamefont {Le}},\ }\href {\doibase
  10.1051/epjconf/20158303007} {\bibfield  {journal} {\bibinfo  {journal} {EPJ
  Web of Conferences}\ }\textbf {\bibinfo {volume} {83}},\ \bibinfo {pages}
  {03007} (\bibinfo {year} {2015})}\BibitemShut {NoStop}%
\bibitem [{\citenamefont {Lampen-Kelley}\ \emph {et~al.}(2018)\citenamefont
  {Lampen-Kelley}, \citenamefont {Janssen}, \citenamefont {Andrade},
  \citenamefont {Rachel}, \citenamefont {Yan}, \citenamefont {Balz},
  \citenamefont {Mandrus}, \citenamefont {Nagler},\ and\ \citenamefont
  {Vojta}}]{Lam18_2}%
  \BibitemOpen
  \bibfield  {author} {\bibinfo {author} {\bibfnamefont {P.}~\bibnamefont
  {Lampen-Kelley}}, \bibinfo {author} {\bibfnamefont {L.}~\bibnamefont
  {Janssen}}, \bibinfo {author} {\bibfnamefont {E.~C.}\ \bibnamefont
  {Andrade}}, \bibinfo {author} {\bibfnamefont {S.}~\bibnamefont {Rachel}},
  \bibinfo {author} {\bibfnamefont {J.-Q.}\ \bibnamefont {Yan}}, \bibinfo
  {author} {\bibfnamefont {C.}~\bibnamefont {Balz}}, \bibinfo {author}
  {\bibfnamefont {D.~G.}\ \bibnamefont {Mandrus}}, \bibinfo {author}
  {\bibfnamefont {S.~E.}\ \bibnamefont {Nagler}}, \ and\ \bibinfo {author}
  {\bibfnamefont {M.}~\bibnamefont {Vojta}},\ }\href
  {https://arxiv.org/abs/1807.06192} {\bibfield  {journal} {\bibinfo  {journal}
  {arXiv:1807.06192}\ } (\bibinfo {year} {2018})}\BibitemShut {NoStop}%
\bibitem [{\citenamefont {Coldea}\ \emph {et~al.}(2002)\citenamefont {Coldea},
  \citenamefont {Tennant}, \citenamefont {Habicht}, \citenamefont {Smeibidl},
  \citenamefont {Wolters},\ and\ \citenamefont {Tylczynski}}]{Col02}%
  \BibitemOpen
  \bibfield  {author} {\bibinfo {author} {\bibfnamefont {R.}~\bibnamefont
  {Coldea}}, \bibinfo {author} {\bibfnamefont {D.~A.}\ \bibnamefont {Tennant}},
  \bibinfo {author} {\bibfnamefont {K.}~\bibnamefont {Habicht}}, \bibinfo
  {author} {\bibfnamefont {P.}~\bibnamefont {Smeibidl}}, \bibinfo {author}
  {\bibfnamefont {C.}~\bibnamefont {Wolters}}, \ and\ \bibinfo {author}
  {\bibfnamefont {Z.}~\bibnamefont {Tylczynski}},\ }\href {\doibase
  10.1103/PhysRevLett.88.137203} {\bibfield  {journal} {\bibinfo  {journal}
  {Phys. Rev. Lett.}\ }\textbf {\bibinfo {volume} {88}},\ \bibinfo {pages}
  {137203} (\bibinfo {year} {2002})}\BibitemShut {NoStop}%
\bibitem [{\citenamefont {Satija}\ \emph {et~al.}(1980)\citenamefont {Satija},
  \citenamefont {Axe}, \citenamefont {Shirane}, \citenamefont {Yoshizawa},\
  and\ \citenamefont {Hirakawa}}]{Sat80}%
  \BibitemOpen
  \bibfield  {author} {\bibinfo {author} {\bibfnamefont {S.~K.}\ \bibnamefont
  {Satija}}, \bibinfo {author} {\bibfnamefont {J.~D.}\ \bibnamefont {Axe}},
  \bibinfo {author} {\bibfnamefont {G.}~\bibnamefont {Shirane}}, \bibinfo
  {author} {\bibfnamefont {H.}~\bibnamefont {Yoshizawa}}, \ and\ \bibinfo
  {author} {\bibfnamefont {K.}~\bibnamefont {Hirakawa}},\ }\href {\doibase
  10.1103/PhysRevB.21.2001} {\bibfield  {journal} {\bibinfo  {journal} {Phys.
  Rev. B}\ }\textbf {\bibinfo {volume} {21}},\ \bibinfo {pages} {2001}
  (\bibinfo {year} {1980})}\BibitemShut {NoStop}%
\bibitem [{\citenamefont {Hentrich}\ \emph {et~al.}(2018)\citenamefont
  {Hentrich}, \citenamefont {Wolter}, \citenamefont {Zotos}, \citenamefont
  {Brenig}, \citenamefont {Nowak}, \citenamefont {Isaeva}, \citenamefont
  {Doert}, \citenamefont {Banerjee}, \citenamefont {Lampen-Kelley},
  \citenamefont {Mandrus}, \citenamefont {Nagler}, \citenamefont {Sears},
  \citenamefont {Kim}, \citenamefont {B\"uchner},\ and\ \citenamefont
  {Hess}}]{Hen18}%
  \BibitemOpen
  \bibfield  {author} {\bibinfo {author} {\bibfnamefont {R.}~\bibnamefont
  {Hentrich}}, \bibinfo {author} {\bibfnamefont {A.~U.~B.}\ \bibnamefont
  {Wolter}}, \bibinfo {author} {\bibfnamefont {X.}~\bibnamefont {Zotos}},
  \bibinfo {author} {\bibfnamefont {W.}~\bibnamefont {Brenig}}, \bibinfo
  {author} {\bibfnamefont {D.}~\bibnamefont {Nowak}}, \bibinfo {author}
  {\bibfnamefont {A.}~\bibnamefont {Isaeva}}, \bibinfo {author} {\bibfnamefont
  {T.}~\bibnamefont {Doert}}, \bibinfo {author} {\bibfnamefont
  {A.}~\bibnamefont {Banerjee}}, \bibinfo {author} {\bibfnamefont
  {P.}~\bibnamefont {Lampen-Kelley}}, \bibinfo {author} {\bibfnamefont {D.~G.}\
  \bibnamefont {Mandrus}}, \bibinfo {author} {\bibfnamefont {S.~E.}\
  \bibnamefont {Nagler}}, \bibinfo {author} {\bibfnamefont {J.}~\bibnamefont
  {Sears}}, \bibinfo {author} {\bibfnamefont {Y.-J.}\ \bibnamefont {Kim}},
  \bibinfo {author} {\bibfnamefont {B.}~\bibnamefont {B\"uchner}}, \ and\
  \bibinfo {author} {\bibfnamefont {C.}~\bibnamefont {Hess}},\ }\href {\doibase
  10.1103/PhysRevLett.120.117204} {\bibfield  {journal} {\bibinfo  {journal}
  {Phys. Rev. Lett.}\ }\textbf {\bibinfo {volume} {120}},\ \bibinfo {pages}
  {117204} (\bibinfo {year} {2018})}\BibitemShut {NoStop}%
\bibitem [{\citenamefont {Knolle}\ \emph {et~al.}(2014)\citenamefont {Knolle},
  \citenamefont {Kovrizhin}, \citenamefont {Chalker},\ and\ \citenamefont
  {Moessner}}]{Kno14}%
  \BibitemOpen
  \bibfield  {author} {\bibinfo {author} {\bibfnamefont {J.}~\bibnamefont
  {Knolle}}, \bibinfo {author} {\bibfnamefont {D.~L.}\ \bibnamefont
  {Kovrizhin}}, \bibinfo {author} {\bibfnamefont {J.~T.}\ \bibnamefont
  {Chalker}}, \ and\ \bibinfo {author} {\bibfnamefont {R.}~\bibnamefont
  {Moessner}},\ }\href {\doibase 10.1103/PhysRevLett.112.207203} {\bibfield
  {journal} {\bibinfo  {journal} {Phys. Rev. Lett.}\ }\textbf {\bibinfo
  {volume} {112}},\ \bibinfo {pages} {207203} (\bibinfo {year}
  {2014})}\BibitemShut {NoStop}%
\bibitem [{\citenamefont {Scheven}\ \emph {et~al.}(1997)\citenamefont
  {Scheven}, \citenamefont {Hannahs}, \citenamefont {Immer},\ and\
  \citenamefont {Chaikin}}]{Sch97}%
  \BibitemOpen
  \bibfield  {author} {\bibinfo {author} {\bibfnamefont {U.~M.}\ \bibnamefont
  {Scheven}}, \bibinfo {author} {\bibfnamefont {S.~T.}\ \bibnamefont
  {Hannahs}}, \bibinfo {author} {\bibfnamefont {C.}~\bibnamefont {Immer}}, \
  and\ \bibinfo {author} {\bibfnamefont {P.~M.}\ \bibnamefont {Chaikin}},\
  }\href {\doibase 10.1103/PhysRevB.56.7804} {\bibfield  {journal} {\bibinfo
  {journal} {Phys. Rev. B}\ }\textbf {\bibinfo {volume} {56}},\ \bibinfo
  {pages} {7804} (\bibinfo {year} {1997})}\BibitemShut {NoStop}%
\bibitem [{\citenamefont {Balz}\ and\ \citenamefont {et~al.}(2019)}]{Bal19}%
  \BibitemOpen
  \bibfield  {author} {\bibinfo {author} {\bibfnamefont {C.}~\bibnamefont
  {Balz}}\ and\ \bibinfo {author} {\bibnamefont {et~al.}},\ }\href@noop {}
  {\bibfield  {journal} {\bibinfo  {journal} {unpublished}\ } (\bibinfo {year}
  {2019})}\BibitemShut {NoStop}%
\bibitem [{Sup()}]{SuppMat}%
  \BibitemOpen
  \href@noop {} {\enquote {\bibinfo {title} {See supplemental material
  for additional details of the magnetocaloric effect measurements}}\
  }\BibitemShut {NoStop}%
\bibitem [{\citenamefont {Zheng}\ \emph {et~al.}(2017)\citenamefont {Zheng},
  \citenamefont {Ran}, \citenamefont {Li}, \citenamefont {Wang}, \citenamefont
  {Wang}, \citenamefont {Liu}, \citenamefont {Liu}, \citenamefont {Normand},
  \citenamefont {Wen},\ and\ \citenamefont {Yu}}]{Zhe17}%
  \BibitemOpen
  \bibfield  {author} {\bibinfo {author} {\bibfnamefont {J.}~\bibnamefont
  {Zheng}}, \bibinfo {author} {\bibfnamefont {K.}~\bibnamefont {Ran}}, \bibinfo
  {author} {\bibfnamefont {T.}~\bibnamefont {Li}}, \bibinfo {author}
  {\bibfnamefont {J.}~\bibnamefont {Wang}}, \bibinfo {author} {\bibfnamefont
  {P.}~\bibnamefont {Wang}}, \bibinfo {author} {\bibfnamefont {B.}~\bibnamefont
  {Liu}}, \bibinfo {author} {\bibfnamefont {Z.-X.}\ \bibnamefont {Liu}},
  \bibinfo {author} {\bibfnamefont {B.}~\bibnamefont {Normand}}, \bibinfo
  {author} {\bibfnamefont {J.}~\bibnamefont {Wen}}, \ and\ \bibinfo {author}
  {\bibfnamefont {W.}~\bibnamefont {Yu}},\ }\href {\doibase
  10.1103/PhysRevLett.119.227208} {\bibfield  {journal} {\bibinfo  {journal}
  {Phys. Rev. Lett.}\ }\textbf {\bibinfo {volume} {119}},\ \bibinfo {pages}
  {227208} (\bibinfo {year} {2017})}\BibitemShut {NoStop}%
\bibitem [{\citenamefont {Sears}\ \emph {et~al.}(2017)\citenamefont {Sears},
  \citenamefont {Zhao}, \citenamefont {Xu}, \citenamefont {Lynn},\ and\
  \citenamefont {Kim}}]{Sea17}%
  \BibitemOpen
  \bibfield  {author} {\bibinfo {author} {\bibfnamefont {J.~A.}\ \bibnamefont
  {Sears}}, \bibinfo {author} {\bibfnamefont {Y.}~\bibnamefont {Zhao}},
  \bibinfo {author} {\bibfnamefont {Z.}~\bibnamefont {Xu}}, \bibinfo {author}
  {\bibfnamefont {J.~W.}\ \bibnamefont {Lynn}}, \ and\ \bibinfo {author}
  {\bibfnamefont {Y.-J.}\ \bibnamefont {Kim}},\ }\href {\doibase
  10.1103/PhysRevB.95.180411} {\bibfield  {journal} {\bibinfo  {journal} {Phys.
  Rev. B}\ }\textbf {\bibinfo {volume} {95}},\ \bibinfo {pages} {180411(R)}
  (\bibinfo {year} {2017})}\BibitemShut {NoStop}%
\bibitem [{\citenamefont {Wolter}\ \emph {et~al.}(2017)\citenamefont {Wolter},
  \citenamefont {Corredor}, \citenamefont {Janssen}, \citenamefont {Nenkov},
  \citenamefont {Sch\"onecker}, \citenamefont {Do}, \citenamefont {Choi},
  \citenamefont {Albrecht}, \citenamefont {Hunger}, \citenamefont {Doert},
  \citenamefont {Vojta},\ and\ \citenamefont {B\"uchner}}]{Wol17}%
  \BibitemOpen
  \bibfield  {author} {\bibinfo {author} {\bibfnamefont {A.~U.~B.}\
  \bibnamefont {Wolter}}, \bibinfo {author} {\bibfnamefont {L.~T.}\
  \bibnamefont {Corredor}}, \bibinfo {author} {\bibfnamefont {L.}~\bibnamefont
  {Janssen}}, \bibinfo {author} {\bibfnamefont {K.}~\bibnamefont {Nenkov}},
  \bibinfo {author} {\bibfnamefont {S.}~\bibnamefont {Sch\"onecker}}, \bibinfo
  {author} {\bibfnamefont {S.-H.}\ \bibnamefont {Do}}, \bibinfo {author}
  {\bibfnamefont {K.-Y.}\ \bibnamefont {Choi}}, \bibinfo {author}
  {\bibfnamefont {R.}~\bibnamefont {Albrecht}}, \bibinfo {author}
  {\bibfnamefont {J.}~\bibnamefont {Hunger}}, \bibinfo {author} {\bibfnamefont
  {T.}~\bibnamefont {Doert}}, \bibinfo {author} {\bibfnamefont
  {M.}~\bibnamefont {Vojta}}, \ and\ \bibinfo {author} {\bibfnamefont
  {B.}~\bibnamefont {B\"uchner}},\ }\href {\doibase 10.1103/PhysRevB.96.041405}
  {\bibfield  {journal} {\bibinfo  {journal} {Phys. Rev. B}\ }\textbf {\bibinfo
  {volume} {96}},\ \bibinfo {pages} {041405(R)} (\bibinfo {year}
  {2017})}\BibitemShut {NoStop}%
\bibitem [{\citenamefont {Baek}\ \emph {et~al.}(2017)\citenamefont {Baek},
  \citenamefont {Do}, \citenamefont {Choi}, \citenamefont {Kwon}, \citenamefont
  {Wolter}, \citenamefont {Nishimoto}, \citenamefont {van~den Brink},\ and\
  \citenamefont {B\"uchner}}]{Bae17}%
  \BibitemOpen
  \bibfield  {author} {\bibinfo {author} {\bibfnamefont {S.-H.}\ \bibnamefont
  {Baek}}, \bibinfo {author} {\bibfnamefont {S.-H.}\ \bibnamefont {Do}},
  \bibinfo {author} {\bibfnamefont {K.-Y.}\ \bibnamefont {Choi}}, \bibinfo
  {author} {\bibfnamefont {Y.~S.}\ \bibnamefont {Kwon}}, \bibinfo {author}
  {\bibfnamefont {A.~U.~B.}\ \bibnamefont {Wolter}}, \bibinfo {author}
  {\bibfnamefont {S.}~\bibnamefont {Nishimoto}}, \bibinfo {author}
  {\bibfnamefont {J.}~\bibnamefont {van~den Brink}}, \ and\ \bibinfo {author}
  {\bibfnamefont {B.}~\bibnamefont {B\"uchner}},\ }\href {\doibase
  10.1103/PhysRevLett.119.037201} {\bibfield  {journal} {\bibinfo  {journal}
  {Phys. Rev. Lett.}\ }\textbf {\bibinfo {volume} {119}},\ \bibinfo {pages}
  {037201} (\bibinfo {year} {2017})}\BibitemShut {NoStop}%
\bibitem [{\citenamefont {Fortune}\ \emph {et~al.}(2009)\citenamefont
  {Fortune}, \citenamefont {Hannahs}, \citenamefont {Yoshida}, \citenamefont
  {Sherline}, \citenamefont {Ono}, \citenamefont {Tanaka},\ and\ \citenamefont
  {Takano}}]{For09}%
  \BibitemOpen
  \bibfield  {author} {\bibinfo {author} {\bibfnamefont {N.~A.}\ \bibnamefont
  {Fortune}}, \bibinfo {author} {\bibfnamefont {S.~T.}\ \bibnamefont
  {Hannahs}}, \bibinfo {author} {\bibfnamefont {Y.}~\bibnamefont {Yoshida}},
  \bibinfo {author} {\bibfnamefont {T.~E.}\ \bibnamefont {Sherline}}, \bibinfo
  {author} {\bibfnamefont {T.}~\bibnamefont {Ono}}, \bibinfo {author}
  {\bibfnamefont {H.}~\bibnamefont {Tanaka}}, \ and\ \bibinfo {author}
  {\bibfnamefont {Y.}~\bibnamefont {Takano}},\ }\href {\doibase
  10.1103/PhysRevLett.102.257201} {\bibfield  {journal} {\bibinfo  {journal}
  {Phys. Rev. Lett.}\ }\textbf {\bibinfo {volume} {102}},\ \bibinfo {pages}
  {257201} (\bibinfo {year} {2009})}\BibitemShut {NoStop}%
\bibitem [{\citenamefont {Winter}\ \emph {et~al.}(2018)\citenamefont {Winter},
  \citenamefont {Riedl}, \citenamefont {Kaib}, \citenamefont {Coldea},\ and\
  \citenamefont {Valenti}}]{Win18}%
  \BibitemOpen
  \bibfield  {author} {\bibinfo {author} {\bibfnamefont {S.~M.}\ \bibnamefont
  {Winter}}, \bibinfo {author} {\bibfnamefont {K.}~\bibnamefont {Riedl}},
  \bibinfo {author} {\bibfnamefont {D.}~\bibnamefont {Kaib}}, \bibinfo {author}
  {\bibfnamefont {R.}~\bibnamefont {Coldea}}, \ and\ \bibinfo {author}
  {\bibfnamefont {R.}~\bibnamefont {Valenti}},\ }\href {\doibase
  10.1103/PhysRevLett.120.077203} {\bibfield  {journal} {\bibinfo  {journal}
  {Phys. Rev. Lett.}\ }\textbf {\bibinfo {volume} {120}},\ \bibinfo {pages}
  {077203} (\bibinfo {year} {2018})}\BibitemShut {NoStop}%
\bibitem [{\citenamefont {McClarty}\ \emph {et~al.}(2018)\citenamefont
  {McClarty}, \citenamefont {Dong}, \citenamefont {Gohlke}, \citenamefont
  {Rau}, \citenamefont {Pollmann}, \citenamefont {Moessner},\ and\
  \citenamefont {Penc}}]{McC18}%
  \BibitemOpen
  \bibfield  {author} {\bibinfo {author} {\bibfnamefont {P.~A.}\ \bibnamefont
  {McClarty}}, \bibinfo {author} {\bibfnamefont {X.-Y.}\ \bibnamefont {Dong}},
  \bibinfo {author} {\bibfnamefont {M.}~\bibnamefont {Gohlke}}, \bibinfo
  {author} {\bibfnamefont {J.~G.}\ \bibnamefont {Rau}}, \bibinfo {author}
  {\bibfnamefont {F.}~\bibnamefont {Pollmann}}, \bibinfo {author}
  {\bibfnamefont {R.}~\bibnamefont {Moessner}}, \ and\ \bibinfo {author}
  {\bibfnamefont {K.}~\bibnamefont {Penc}},\ }\href {\doibase
  10.1103/PhysRevB.98.060404} {\bibfield  {journal} {\bibinfo  {journal} {Phys.
  Rev. B}\ }\textbf {\bibinfo {volume} {98}},\ \bibinfo {pages} {060404(R)}
  (\bibinfo {year} {2018})}\BibitemShut {NoStop}%
\end{thebibliography}
\end{document}